\documentclass[iop,twocolumn,twocolappendix,numberedappendix]{emulateapj}
\usepackage{amsmath}
\usepackage{times}
\usepackage{graphicx}
\usepackage{xcolor}
\usepackage{bm}
\usepackage{hyperref}
\hypersetup{colorlinks = true, linkcolor = blue, anchorcolor = red, 
     citecolor = blue, filecolor = red, urlcolor = blue, backref = page}

\newcommand{\MR}{{\cal R}}

\newcommand{\Msun}{M_{\odot}}

\begin{document}
\title{Adding a Suite of Chemical Abundances to the MACER 
        Code \\for the Evolution of Massive Elliptical Galaxies}
        
\author{Zhaoming Gan$^{1}$\footnotemark[$$], Ena Choi$^2$,   
            Jeremiah P. Ostriker$^{2}$\footnotemark[$$], Luca Ciotti$^3$,
            Silvia Pellegrini$^3$}
\affil{$^{1}$Shanghai Astronomical Observatory, Chinese Academy of Sciences,
        80 Nandan Road, Shanghai 200030, China}
\affil{$^{2}$Department of Astronomy, Columbia University,
        550 W, 120th Street, New York, NY 10027, USA}
\affil{$^{3}$Department of Physics and Astronomy, University of Bologna,
        via Piero Gobetti 93/2, 40129 Bologna, Italy}

\begin{abstract}
We add a suite of chemical abundances to the \texttt{MACER} 
(Massive AGN Controlled Ellipticals Resolved) 2D code, 
by solving 12 additional continuity equations for
H, He, C, N, O, Ne, Mg, Si, S, Ca, Fe and Ni respectively 
with sources from AGB stars and supernovae of type Ia and II
with metal yields based on standard stellar physics. 
New stars, formed in Toomre unstable circumnuclear disks 
(of a size $\lesssim150$ parsec), are 
assumed to have a top-heavy initial mass function with  a power index of 1.65. 
The metal dilution effects due to cosmic accretion are also included. 
With the high resolution of few parsecs in central regions, 
resolved black hole accretion and 
AGN feedback,  we can track the metal enrichment, 
transportation and dilution throughout the modeled massive elliptical galaxy of velocity dispersion 
$\sim280$ km/s. We retrieve the chemical composition of 
the BAL winds launched by the central AGN, 
synthesize the X-ray features of the hot ISM,
and find that 
(1) the simulated metallicity  in the BAL winds could be up to $\sim 8 Z_\odot$, 
while that of the hot ISM in the host galaxy is $\sim 2.3 Z_\odot$, matching well with 
SDSS observations of BLR gas;
(2) the X-ray emitting hot gas is metal enriched with a typical value $\sim2.5 Z_\odot$;
(3) the circumunuclear cold gas disk,
where the metals are condensed, further enriched and recycled,
plays a critical role in the metal enrichment;
(4) the black hole accretion rate $\dot{M}_{\rm BH}$ linearly correlates with
the star formation rate $\dot{M}_\star^+$ in the circumnuclear disk, 
i.e, $\dot{M}_\star^+ \sim 7.7 \dot{M}_{\rm BH}$, but lagged in time by roughly $10^6$ years.
\end{abstract}


\section{Introduction} \label{sec:introduction}
We have developed the \texttt{MACER} (Massive AGN Controlled Ellipticals Resolved) code 
over some time as an instrument for exploring the evolution of massive elliptical galaxies 
at high spatial resolution including a relatively complete set of physical processes. 
A recent paper \citep{gan_macer_2018} outlines the details of how rotation, 
massive dark matter halos, infall of cosmological gas and other features were 
recently added and tested. With this additional physical infrastructure 
we discovered features expected in the normal evolution of elliptical galaxies 
that surprised us but for which there is ample observational evidence: 
specifically the formation of circumnuclear cold gaseous discs that are Toomre unstable 
and which form massive stars in episodic bursts that fuel the central black hole and 
explain the enigmatic ``E+A'' phenomenon. The code, with resolution of parsecs 
in the central region, resolves the {fiducial} Bondi radius and hence can treat black hole accretion 
and AGN (Active Galactic Nucleus) feedback 
in some detail including both radiative (UV \& X-ray) and 
BAL (Broad Absorption Line) winds in a fashion that imitates the observed output of black holes 
in both their low and high output states.

In this {introductory} paper we describe how we can add 12 chemical species produced 
by standard stellar physics and injected into the interstellar medium (ISM) by 
asymptotic branch stars (AGBs), type I and type II supernovae (SNe I, SNe II). 
This will enable us to address a whole new range of questions. For example, 
``What is the chemical composition expected of the BAL winds?'' or 
``What iron abundance should be seen in the hot X-ray emitting gas?''. 
These are questions that most cosmologically based galaxy evolution codes would have 
difficulty addressing because they typically do not have the spatial resolution 
{needed to resolve} the central regions where the winds are generated 
and a significant part of the new metals are produced.

Half of the observed elliptical galaxies are known \textcolor{black}{to contain} 
cold gaseous discs in their centers
(a.k.a circumnuclear disks, e.g., 
\citealt{sarzi_sauron_2006, davis_atlas3d_2011, boizelle_alma_2017}). 
As we demonstrated in \citet{gan_macer_2018}, 
the formation of circumnuclear disk is inevitable
due to the angular momentum barrier, as rotation is the general case. 
Since the length scale of an AGN is very small when compared to the galactic scale,
most of the infalling gas (due to radiative cooling) 
will {settle in a flat, rotationally supported disk-like structure} 
before it reaches the galaxy center (see also \citealt{eisenreich_active_2017}).
Moreover, the gas will cool down further in the circumnuclear disk,
which makes it ideal for star formation.

As cold gas condenses and accumulates on the circumnuclear disk,
such a cold gaseous disk would become highly over-dense, and we found that 
some of the over-dense disk rings would be gravitationally unstable 
(a.k.a. the Toomre Instability; \citealt{toomre_gravitational_1964, tan_star-forming_2005}).
Consequently, angular momentum transfer and star formation are allowed 
at the same time due to the Toomre instability
\citep[e.g.,][]{gammie_nonlinear_2001,goodman_self-gravity_2003, 
rice_investigating_2005,thompson_radiation_2005, machida_formation_2010}. 
Therefore, one may expect a near coincidence of star formation and AGN activity 
--- the former is an important metal enriching source 
(by SNe II; \citealt{goodman_supermassive_2004}), 
and via accretion onto the supermassive black hole from the circumnuclear disk 
and then the AGN wind feedback, 
some of the metal enriched gas {will} be recycled
in form of BAL winds --- 
in this paper, we will also demonstrate the key role of the circumnuclear disks
in the chemical evolution of massive elliptical galaxies, i.e.,
where the metals are condensed, further enriched and recycled.
	 
In the next section 
of this paper we will provide the details of 
how we implement the chemical abundance inputs to the code. 
We outline briefly in \S\ref{sec:model-improvement} the model improvements
we have made to facilitate the introduction of diverse metallicity components.
In \S\ref{sec:results} we will present some results of computations, 
including the chemical distribution in the modeled galaxy, 
the radiative features of the metal rich gas, composition expected in the BAL winds. 
Conclusions are reserved for final sections.

\section{Chemical Abundances} \label{sec:model}
In the \texttt{MACER} simulations, we solve the continuity equation of the ISM, 
together with the conservation laws of momentum and energy, 
including source terms due to the stellar evolution of
AGBs  ($\dot{\rho}_{\star} $), 
SNe Ia ($\dot{\rho}_{\rm I} $) and
SNe II ($\dot{\rho}_{\rm II}$) 
(\citealt{ciotti_cooling_1997, novak_feedback_2011, gan_macer_2018}), 
i.e., 
\begin{equation} \label{eq:massconsvr}
   \frac{\partial \rho}{\partial t} + \nabla\cdot(\rho{\bf v}) + \nabla\cdot\dot{\bf m}_{\rm Q}  
   = \dot{\rho}_{\star} + \dot{\rho}_{\rm I} + \dot{\rho}_{\rm II} - \dot{\rho}_{\star}^{+},
\end{equation}
where  $\rho$ and ${\bf v}$ are the mass density and fluid velocity, respectively.
$\nabla\cdot\dot{\bf m}_{\rm Q}$ and $\dot{\rho}_{\rm \star}^{+}$ are the mass advection 
due to the Toomre instability \citep{toomre_gravitational_1964} 
and the mass sink term due to star formation, respectively.
We refer to our code paper (\citealt{gan_macer_2018} and references therein) 
for the full description of the physics and equations we solve in the simulations. 

In this paper, we intend to track the chemical evolution of the metals 
by using 12 tracers 
$X_i$ (i=1, 2, ..., 12; {mass of each element per unit volume}) 
for H, He, C, N, O, Ne, Mg, Si, S, Ca, Fe and Ni, respectively 
(see also \citealt{eisenreich_active_2017}, \citealt{choi_physics_2017} 
for large scale simulations using SPH).
We solve 12 additional continuity equations of the tracers, assuming the chemical species 
co-move once after they are injected into the ISM, i.e.,
\begin{equation} \label{eq:metal-tracers}
   \frac{\partial X_{\rm i}}{\partial t} + \nabla\cdot(X_{\rm i}{\bf v}) + \nabla\cdot\dot{\bf m}_{\rm Q,i} 
        = \dot{X}_{\rm \star,i} + \dot{X}_{\rm I,i} + \dot{X}_{\rm II,i} - \dot{X}_{\rm \star,i}^{+},
\end{equation}
where
\begin{equation} 
   \dot{\bf m}_{\rm Q,i} = (X_{\rm i}/\rho) \cdot \dot{\bf m}_{\rm Q}, \quad
   \dot{X}_{\rm \star,i}^{+} = (X_{\rm i}/\rho) \cdot \dot{\rho}^{+}_{\star},
\end{equation}
and ${\bf v}$ is obtained by solving the hydrodynamical equations as ususal.
In Equation \ref{eq:metal-tracers}, the passive stellar evolution, i.e., AGBs ($\dot{X}_{\rm \star,i}$), 
SNe Ia ($\dot{X}_{\rm I,i}$) and SNe II ($\dot{X}_{\rm II,i}$),  serves as metal-enriching sources, 
while with different metal compositions (see Table \ref{tab:metal}).  
The advection terms $\nabla\cdot(X_{\rm i}{\bf v})$ and $\nabla\cdot\dot{\bf m}_{\rm Q,i}$ 
\textcolor{black}{describe the transport and the mixing of} ISM with different metal abundances.
Star formation ($\dot{X}_{\rm \star,i}^{+}$) is treated as a sink of local metals, 
but not to change the {local} abundance.

\begin{table}[ht]
\caption{Mass fraction of the elements from various sources}\label{tab:metal}
\begin{center}
\begin{tabular}{lccccc}
\hline\hline
{  } & {AGBs$^a$} &{SNe Ia$^b$} & {SNe II$^c$} & {CGM$^d$} & {Solar$^e$} \\
\hline
H &  0.71287 &  0.00000 &  0.51800 &  0.74682 &  0.73810 \\
He&  0.26702 &  0.00000 &  0.33483 &  0.25117 &  0.24850 \\
C &  0.00294 &  0.00225 &  0.01031 &  0.00036 &  0.00241 \\
N &  0.00183 &  0.00000 &  0.00362 &  0.00011 &  0.00071 \\
O &  0.00872 &  0.07465 &  0.08175 &  0.00087 &  0.00585 \\
Ne&  0.00190 &  0.00264 &  0.02518 &  0.00019 &  0.00127 \\
Mg&  0.00106 &  0.01123 &  0.00761 &  0.00011 &  0.00071 \\
Si&  0.00047 &  0.21212 &  0.00391 &  0.00010 &  0.00068 \\
S &  0.00101 &  0.08500 &  0.00861 &  0.00005 &  0.00031 \\
Ca&  0.00010 &  0.01086 &  0.00050 &  0.00001 &  0.00007 \\
Fe&  0.00197 &  0.54693 &  0.00540 &  0.00020 &  0.00132 \\
Ni&  0.00011 &  0.05432 &  0.00028 &  0.00001 &  0.00007 \\
\hline
 Z&   0.0201 &   1.0000 &   0.1472 &   0.0020 &   0.0134 \\
\hline \hline
\end{tabular}
\end{center}
\hangindent 0.75em
$^a$ averaged metal abundance of AGB winds over the time span from $t_{\rm age} = 2$
         to 13.7 Gyr, i.e., $(<{\dot{X}_{\rm \star, i}}/\dot{\rho}_{\rm \star}>)$ 
         \citep{karakas_updated_2010}; 
         
$^b$ metal abundance of SNe Ia ejecta, i.e., {$(\dot{X}_{\rm I, i}/\dot{\rho}_{\rm I})$} 
	\citep{seitenzahl_three-dimensional_2013};
	
$^c$ metal abundance of SNe II ejecta, i.e., {$(\dot{X}_{\rm II, i}/\dot{\rho}_{\rm II})$} 
	\citep{nomoto_nucleosynthesis_2013};
	
\hangindent 0.75em	
$^d$ metal abundance of the low-metallicity infalling CGM  which is made of 1/4 of primordial gas
        and 3/4 low metallicity gas of 0.2 solar abundance; 
        
$^e$ solar abundance \citep{asplund_chemical_2009};

$^f$ metallicity Z, i.e, mass fraction of all chemical species except H and He.
\end{table}

We start the simulations with an initial stellar population of an age of 2 Gyr, 
its mass distribution is determined by a galaxy dynamics model 
(\citealt{gan_macer_2018}).
As cooling flows develop, a circumunuclear disk forms naturally in the galaxy
center, in which the gas is cold and over-dense. Star formation is then inevitable. 
Therefore, we have two stellar populations in our modeled galaxy:
(1) the initial stellar population which is 2 Gyr old at the beginning of the simulation, 
and in which all massive stars have died, 
while its secular evolution (i.e., AGBs and SNe Ia) is considered in our model.
(2) a new stellar population actively forming
in the central cold gas circumnuclear disk during the simulations
and producing SNe II that plays an important role in the metal enrichment.

We calculate time-dependent nucleosynthesis output returned
to the ISM by evolving stars of a simple stellar population (SSP)
with ``CELib'', an open-source software library for chemical evolution 
\citep{saitoh_chemical_2017}.

\begin{figure}[htb] 
\centering
\includegraphics[width=0.45\textwidth]{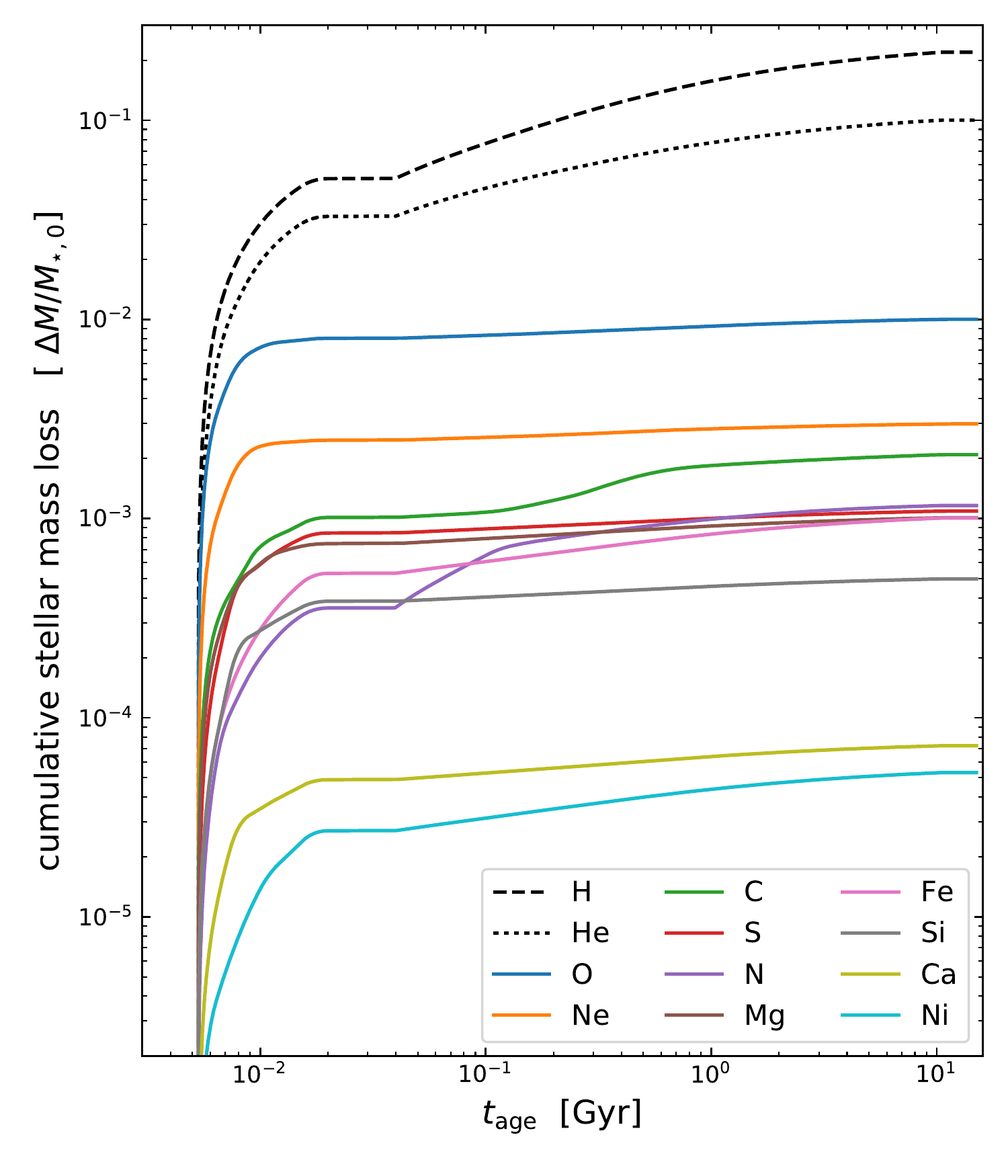}
\caption{Cumulative mass loss from the initial stellar population in individual chemical species. 
The lines in colors are the mass fractions of the metals ejected during the passive stellar evolution.
The mass losses are in units of $M_{\star,0}$ (i.e.,
the total mass of the initial stellar population at $t_{\rm age}=0$.)}
\label{fig:metal}
\end{figure}

For the initial stellar population, 
we assume a SSP that consists of stars of identical age and chemical
composition, and follows the \citet{kroupa_variation_2001} initial
mass function (IMF) with a stellar mass range of $0.1 \-- 120 \Msun$. 
The SSP is assumed of metallicity $Z=1.5Z_\odot$ 
(where $Z_\odot=0.0134$ is the solar metallicity as quoted in Table \ref{tab:metal}).
This abundance matches that expected for an elliptical as massive 
as the modeled one, which should be supersolar  (e.g., \citealt{thomas_environment_2010}).
The metallicity-dependent stellar lifetime 
is taken from \cite{portinari_galactic_1998}.  The stellar ``yields'', 
the amount of newly synthesized and ejected elements via 
stellar evolution, are adopted from \cite{nomoto_nucleosynthesis_2013}, 
\cite{doherty_super_2014} and \cite{karakas_updated_2010} 
for high mass stars (Type II SNe), massive AGB stars with 
$M>6 \Msun$, and low mass AGB stars respectively. Our 
time-dependent chemical enrichment calculation shows that the 
nucleosynthesis output enriches the ISM mainly via two phases,
an early ($t \lesssim 10$ Myr) phase driven by Type II SNe, and 
a subsequent late phase ($t \gtrsim 40$ Myr) driven by AGB stars 
(cf. Equation \ref{eq:stellar-mass-loss-rate}). 
Since our simulation starts at $t_{\rm age}=2$ Gyr, there will be only AGBs,
and its time-dependent abundance is used in our model as shown in Figure \ref{fig:metal}, 
while the averaged metal abundance of  AGB winds
is also summarized in Table \ref{tab:metal} for the readers' reference.

The initial stellar population also contributes metals via Type Ia SNe 
(cf. Equation \ref{eq:snia-rate}). 
\textcolor{black}{We adopted the ejecta mass and the nucleosynthesis yields of SN Ia from the N100 model of \citet{seitenzahl_three-dimensional_2013}.  According to three-dimensional high resolution hydrodynamical simulations of SN Ia explosion model of \citet{seitenzahl_three-dimensional_2013}, each SNIa explosion distributes 1.35 Msolar ejecta in total, and the chemical composition of the ejecta is Fe, Silicon and Oxygen dominated as summarized in Table \ref{tab:metal}.}

The new stellar population contribute metals, unlike the initial stellar population,
mostly in its early phase evolution via massive stars. 
For simplicity, we only consider the metals ejected by Type II SNe.
For such new stellar population formed by gravitational instability 
in Toomre unstable circumnuclear disks, we adopt
the \citeauthor{kroupa_variation_2001} IMF, 
but with the top-heavy index of 1.65 advocated by 
\citet{bartko_extremely_2010} and \citet{lu_stellar_2013}, 
which is based on the observed stellar disks in the 
MW and M31. We adopt a stellar mass range of $1 \-- 50 \Msun$ 
(cf. Equation \ref{eq:star-formation-imf}).
The time-averaged metal abundance of its SN II phase
is summarized in Table \ref{tab:metal}.

As in \citet{gan_macer_2018}, we also consider the cosmic accretion (i.e., CGM infall) 
of low-metallicity gas (cf. Table \ref{tab:metal}) through the galaxy outskirts, 
which will dilute the metallicity as it falls into the galaxy and mixes with the ISM there.  
In addition, we also recycle the metals via the BAL winds which are injected back 
to the galaxy by the central AGN.
Those two processes do not appear explicitly in the equations above, but are 
implemented via the designed outer and inner boundary conditions, respectively.
In this way, we can track the chemical evolution, i.e., the metal enrichment, 
transportation and dilution, throughout massive elliptical galaxies.
After AGN bursts there are significant mass outflows from the galaxy.

To evaluate the metal abundance of the BAL winds, we keep tracking the chemical composition
in our AGN ``sub-grid'' model \citep{ciotti_cooling_1997, gan_macer_2018}. 
As  the mass source of the AGN fueling is accreted via 
the inner boundary, we record all the metals that pass through to the galaxy center, 
so that we are able to calculate the averaged metal abundance of the black hole accretion disk 
at any given time. The BAL winds are injected (at the inner edge) back to the galaxy 
with the instantaneous metal abundance of the black hole accretion disk. 

Finally, we assume the initial ISM is of the solar abundance  \citep{asplund_chemical_2009}.
We must emphasize that we have not yet corrected for the depletion of metals
onto dust, and thus our characterization of element abundance is the total
(dust plus gas phase) mass in that element per unit volume in units
of the local mass density. Especially in the cold gas component, 
the fractional depletion of refractory elements onto dust grains
may be a large correction \citep{hensley_grain_2014}.  We reserve it to our future work.

\section{Model Improvements} \label{sec:model-improvement}
Compared to \citet{gan_macer_2018}, the model setups have been modified as follows 
(besides adding the metal tracers) to facilitate the metallicity changes, 
to improve the model self-consistency, and to better fit the observed galaxy properties:

I. the initial black hole mass is increased by 
a factor of 2  (i.e., to $M_{\rm BH}=6.7\times10^8 M_\odot$) 
so as to better match the observed $M_{\rm BH}-\sigma$ relation 
\citep{kormendy_coevolution_2013,yuan_active_2018}, 
while the rest of the galaxy model parameters are the same as in \citet{gan_macer_2018}, 
i.e., total stellar mass at the present time $M_\star=3.35\times10^{11}M_\odot$, 
the effective radius $R_e=7$ kpc, the galaxy ellipticity $\eta=0.2$,
and the dimensionless parameters for the total gravity mass $\MR=20$ 
and for the length scale $\xi=20$ of the modeled galaxy, 
which results in a central projected stellar velocity dispersion of $\sim280$ km/s.

II. in order to describe the ordered rotational velocity 
of the stellar component we adopted the usual Satoh decomposition
in our previous papers, with a constant rotation parameter $k$
\citep{ciotti_effect_2017,yoon_active_2018,gan_macer_2018}. 
Here we add some flexibility to the modeling by considering 
a central value ($k_0$) for the k parameter, 
and we reduce it in the outer parts of the galaxy as
\begin{equation} \label{eq:rotation-profile}
     k = k_0 \cdot e^{-r/R_e},
\end{equation}
where we adopt $k_0=0.25$.
In this way we can have some control over the amount 
of angular momentum stored in the external regions of the galaxy.

III. we adopt the stellar (AGB)
mass loss formula in \citet{pellegrini_hot_2012} 
 and adopt the coefficient $A=3.3$ to match the assumption of the \citeauthor{kroupa_variation_2001} IMF, i.e.,
 \begin{equation} \label{eq:stellar-mass-loss-rate}
     \dot{M}_\star = 10^{-12} A \cdot \frac{M_\star}{M_{\odot}}
     \left(\frac{t_{\rm age}}{12~{\rm Gyr}}\right)^{-1.3}  
     \ \  M_\odot/{\rm year},
\end{equation}
where $t_{\rm age}$ is the age of the initial stellar population, 
and  the fitting formula is valid when  $t_{\rm age} \geq 2$ Gyr.
As usual, the SN Ia rate is evaluated as 
 \begin{equation} \label{eq:snia-rate}
     R_{\rm SN} = 0.16\times10^{-12} h^2 \frac{L_{\rm B}}{L_{\odot}}
     \left(\frac{t_{\rm age}}{12~{\rm Gyr}}\right)^{-1.1}  {\rm year}^{-1},
\end{equation}
where $h$ is the Hubble constant in units of $70~{\rm km}~{\rm s}^{-1}{\rm Mpc}^{-1}$, 
and the B-band luminosity $L_{\rm B}$ is derived by adopting a mass-to-light ratio 
$\Gamma\equiv M_\star/L_{\rm B}=5.8$ in units of the solar value,
as appropriate for an old stellar population with a Kroupa IMF 
\citep{pellegrini_hot_2012}.

\begin{figure*}[htb]
\centering
\includegraphics[width=0.275\textwidth]{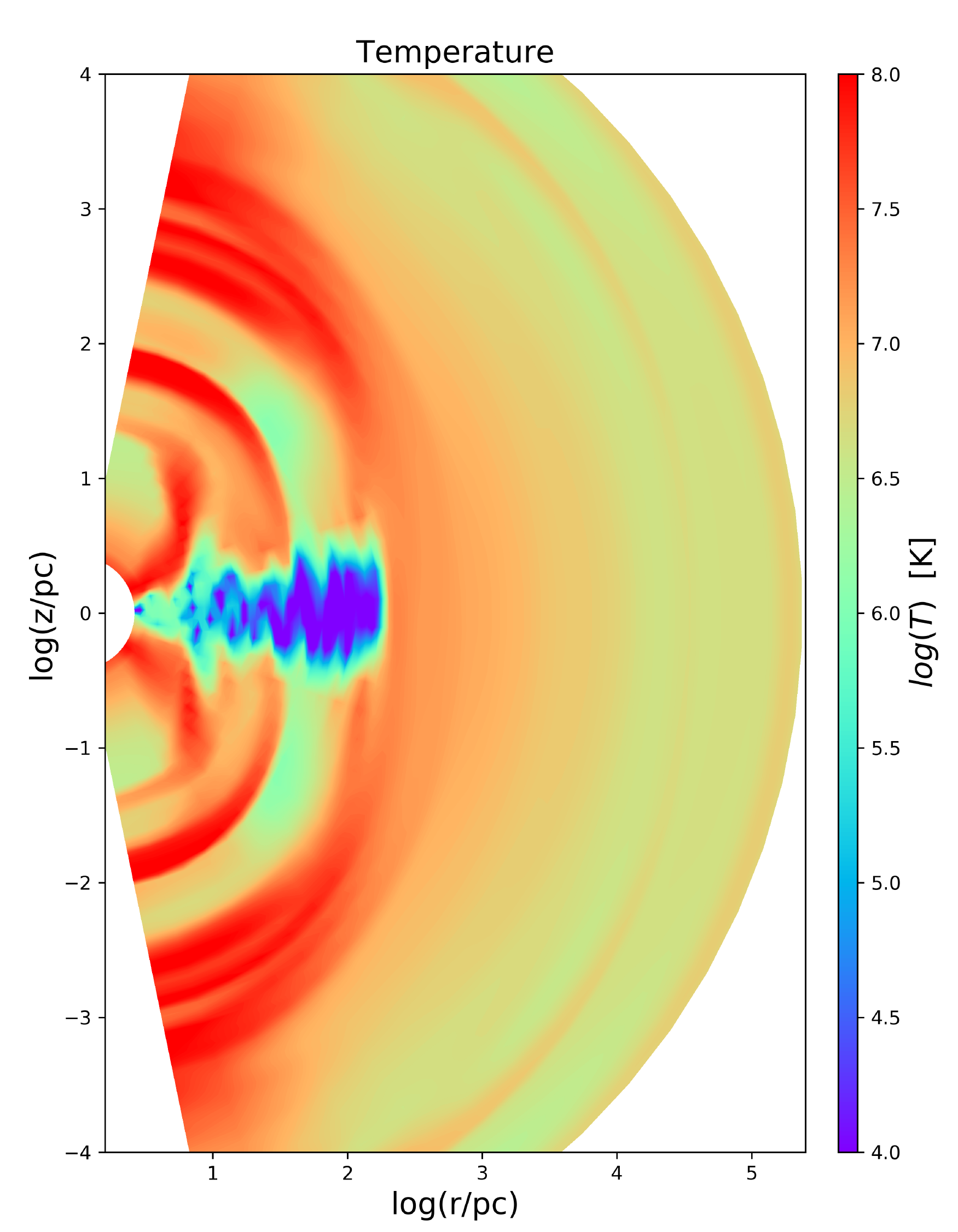}
\includegraphics[width=0.275\textwidth]{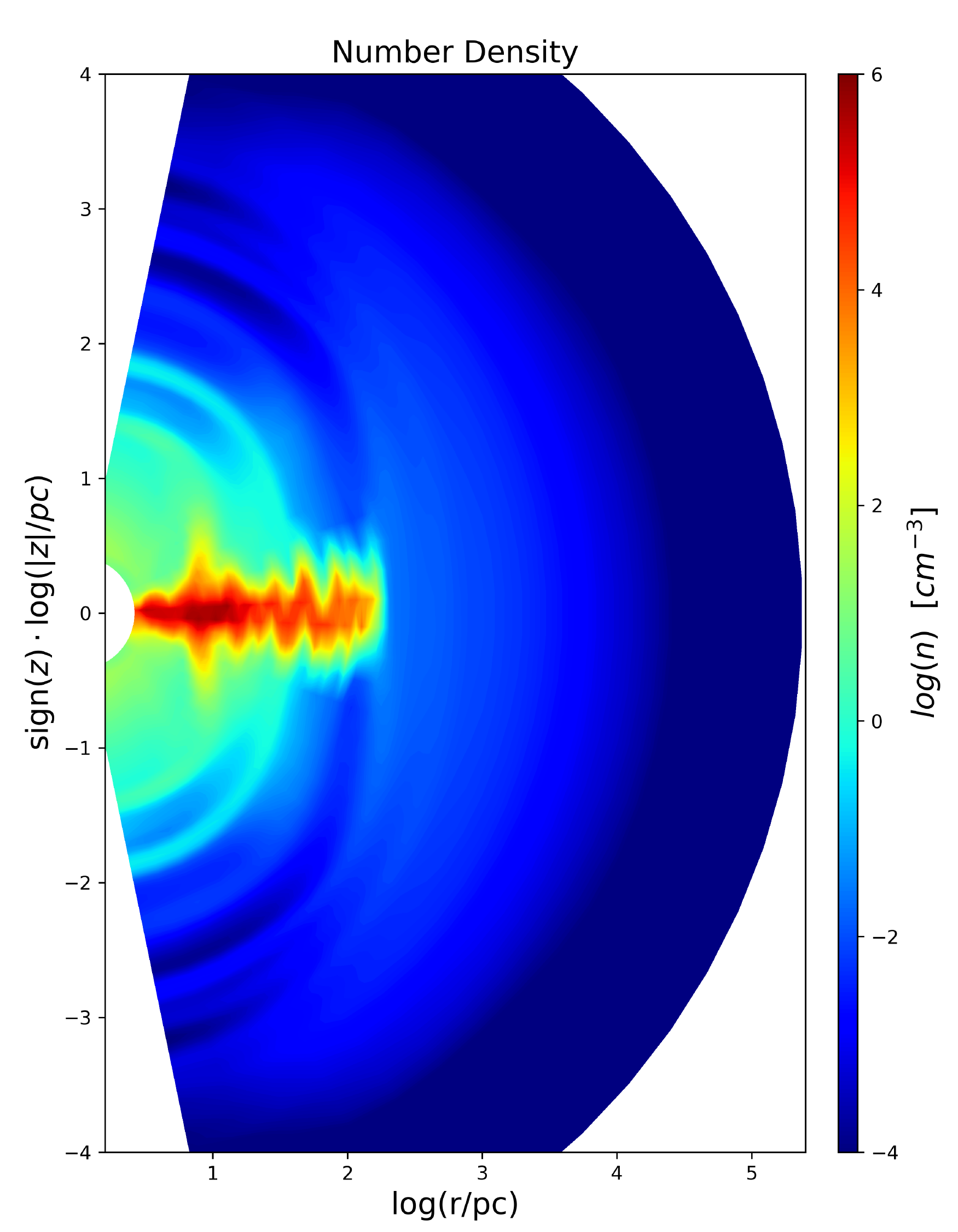}
\includegraphics[width=0.275\textwidth]{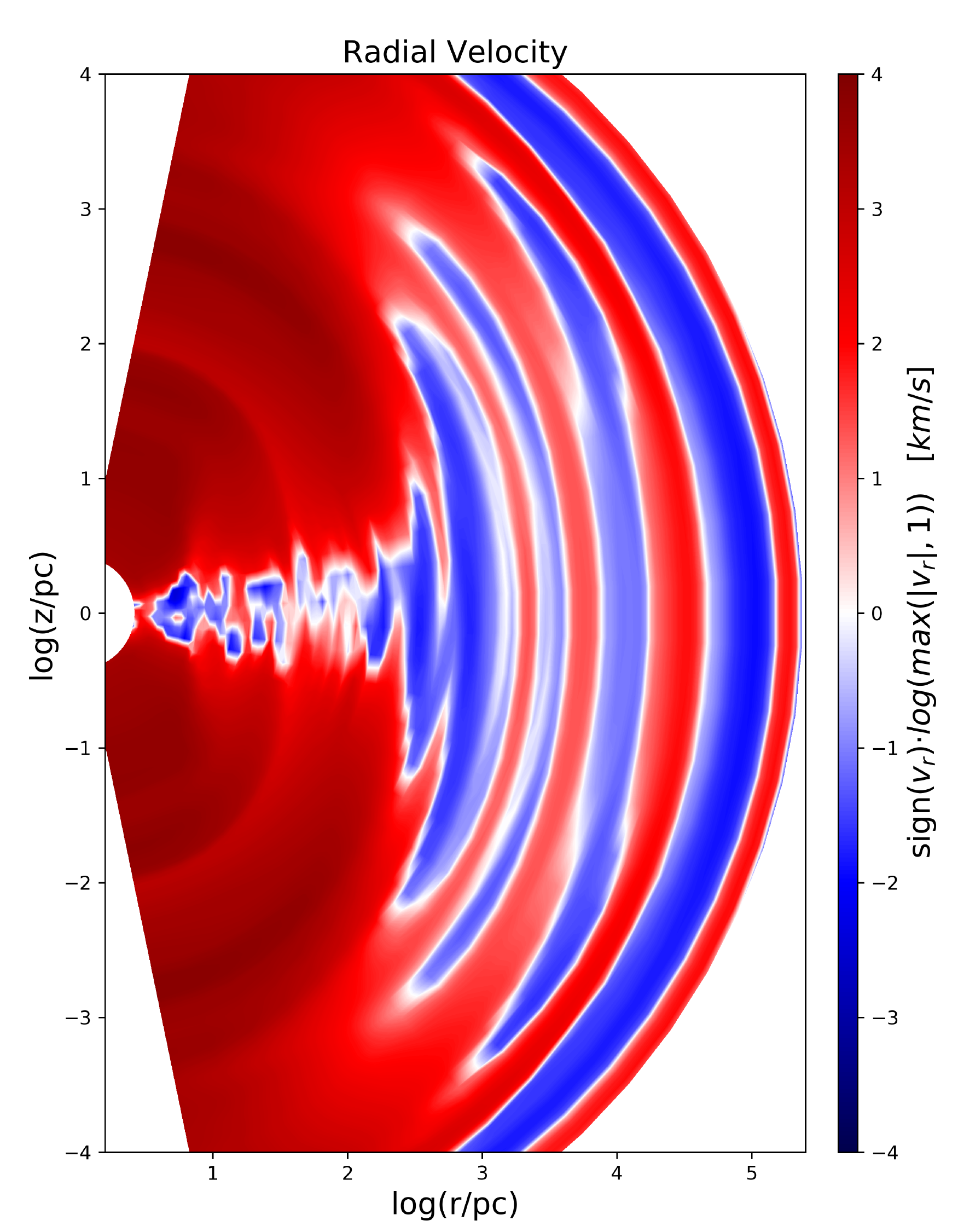}
\caption{Hydrodynamical properties of the ISM during the burst event at $t_{\rm age}=12.1$ Gyr. The size of the cold gaseous disk is $\sim150$ parsec. The high-speed BAL winds is of a bi-conical structure and can be heated up to $10^8$ K. Note the logarithmic radial scale.}
\label{fig:hydro-properties}
\end{figure*}

\textcolor{black}{IV}. the initial mass function of the newly formed stars in the Toomre unstable disks 
is assumed to have a top-heavy profile based on both observations and
theoretical expectations
\citep{goodman_supermassive_2004, bartko_extremely_2010, 
          jiang_star_2011, lu_stellar_2013}, we assume 
\begin{equation} \label{eq:star-formation-imf}
     \frac{dN}{dM} = \frac{N_0}{M_\odot} \left(\frac{M}{M_\odot}\right)^{-1.65}
\end{equation}
with a mass range of $1 \-- 50 \Msun$.
Such an IMF gives that $\sim60\%$ 
of the total new star mass is in massive stars ($M>8M_\odot$), 
which will turn into SNe II in a timescale of $\sim2\times10^7$ year.
We assume each SN II leaves a neutron star of $1.4M_\odot$ and 
ejects $10^{51}$ erg of energy into the ISM (cf. \citealt{ciotti_agn_2012}).

V. we allow for the fact that the atomic heating/cooling function ($S_{\rm line}$)
is approximately proportional to the local metallicity ($Z \equiv 1-(X_1+X_2)/\rho$)
in units of the solar value $Z_\odot = 0.0134$, 
i.e., in the net heating rate $H - C = n^2 (S_{\rm comp} + S_{\rm brem} + S_{\rm line})$,  
\begin{equation} \label{eq:line-recomb}
        S_{\rm line} =  10^{-23}{{a  + b\, (\xi/\xi_0)^c}\over 1 + (\xi/\xi_0)^c} {Z \over Z_\odot},
\end{equation}
$S_{\rm comp}$ and $S_{\rm brem}$ 
are as usual the Compton heating/cooling and the bremsstrahlung loss, respectively
(detailed description of the heating/cooling functions above can be found
in \citealt{ciotti_agn_2012}).

VI. we use the same CGM infall profile as in \citet{gan_macer_2018}, 
however, since we start the simulation when the galaxy is 2 Gyr old, 
we now also start the CGM infall at that time. 
The accumulated mass due to the CGM infall before 
$t_{\rm age} = 2$ Gyr is now taken as the initial remnant ISM,
which is $\sim 10^{10} M_\odot$.

VII. the AGN feedback wind efficiency $\epsilon_w$ is reduced from 0.005 to 0.004.


Finally, we solve the ISM hydrodynamical equations, together with the metal tracers, using the grid-based
\texttt{Athena++} code \citep[version 1.0.0;][]{stone_athena:_2008} in spherical coordinates. 
We assume axi-symmetry but allow rotation (a.k.a. 2.5-dimensional simulation).  
The outer boundary is chosen as 250 kilo-parsec to enclose the whole massive elliptical galaxy, 
the inner boundary $R_{\rm in}$ is set to be 2.5 parsec to resolve the Bondi radius.
We use a logarithmic grid ($\Delta r_{\rm i+1}/ \Delta r_{\rm i} = 1.1$) to 
divide the radial axis into 120 discrete cells. The azimuthal angle $\theta$ 
is divided into 30 uniform cells and covers an azimuthal range from $0.05\pi$ to $0.95\pi$. 
The numerical solver for the gas dynamics is composed 
by the combination of the HLLE Riemann Solver, the PLM reconstruction 
and the second-order van Leer integrator.

\section{Results}\label{sec:results}
Before we address the simulation details, it is helpful to discuss the 
``closed-box problem'', 
i.e., what metal composition could be expected if the metal enrichment is only contributed 
by the secular evolution of the initial old stellar population, i.e., via AGBs and SNe Ia?
We can see from Figure \ref{fig:metal} that the metal composition of AGB winds 
provides only weak evolution after $t_{\rm age}=2$ Gyr (when the simulation starts),
and we use a fixed metal composition for SN Ia ejecta.
Moreover, Equations \ref{eq:stellar-mass-loss-rate} \& \ref{eq:snia-rate} give that
the ratio between the mass return rates of SNe Ia and AGBs evolves weakly 
from 0.008 \textcolor{black}{(at $t_{\rm age}=2$ Gyr)} 
to 0.012 \textcolor{black}{(at $t_{\rm age}=13.7$ Gyr)}. 
Therefore, the secular stellar evolution (AGBs + SNe Ia) 
alone will result in a characteristic metallicity in a narrow range of $2.1 Z_\odot < Z < 2.4Z_\odot$ 
throughout the host galaxy.

In addition, as mentioned in \S\ref{sec:model}, SNe II contribute to the metal enrichment 
when there is star formation,
while star formation mainly occurs in the circumnuclear disk, 
where most of the infalling gas is circularized because of angular momentum barrier.
Ideally, local metallicity could be up to $12.7 Z_\odot$ if SNe II dominate the metal
enrichment (cf Table \ref{tab:metal}). 

We will demonstrate in the rest of this section that the metal enrichment 
in the circumnuclear disk is the key to
understanding the chemical evolution of the modeled galaxy.
Therefore, it is worthwhile to analyze the circumnuclear disk in advance 
(see \S\ref{sec:CND}). 
Following the physical sequence, i.e., 
formation of the circumnuclear disk, star formation, mass inflow, BAL winds, and
then metal transportation, we perform in \S\ref{sec:sf-bha-correlation} detailed analysis 
on the star formation and black hole feeding processes.
In \S\ref{sec:metal-budget}, we present the overall gaseous mass and metal budget.
In \S\ref{sec:metal-distribution}, the spatial distribution of metals is presented, 
and we will see the role of the BAL winds in transporting metals throughout the modeled galaxy.
Finally, the radiative features of the metal-enriched hot ISM are calculated in \S\ref{sec:radiative-featues}.

\begin{figure}[htb]
\centering
\includegraphics[width=0.475\textwidth]{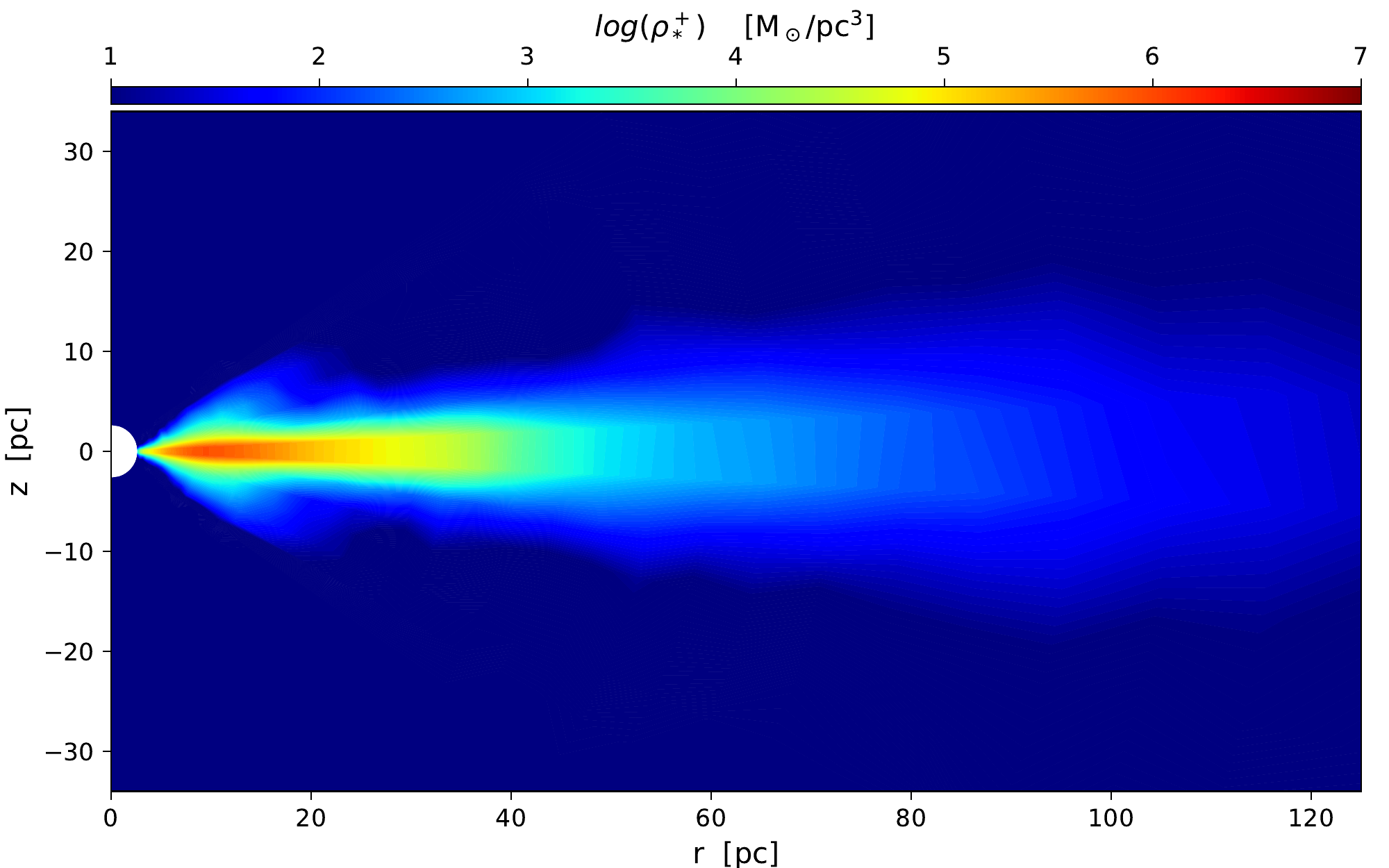}
\caption{Cumulative star formation in the circumnuclear disk at $t_{\rm age}=12.1$ Gyr.  
The size of the stellar disk is $<150$ parsec.
The cumulative star formation in the disk is $\sim 6.5\times10^9 M_\odot$.
The averaged age and metallicity of the new stellar population are 7.52 Gyr 
and $Z/Z_\odot=5.82$, respectively.}
\label{fig:star-formation}
\end{figure}

\subsection{Circumnuclear disk} \label{sec:CND}
In Figure \ref{fig:hydro-properties}, we present the hydrodynamical properties 
of the ISM during the burst event at $t_{\rm age}=12.1$ Gyr. 
The gaseous counterpart of the circumnuclear disk is found
of a size $\sim150$ parsec, and of mass $\sim2.5\times10^8M_\odot$
(see Figure \ref{fig:star-formation} for the stellar counterpart). 
It is shown in the polar regions that the high-speed BAL winds 
launched by the central AGN
can be heated up to $10^8$ K, which should be capable in emitting X rays. 
We have designated as BAL wind gas flowing out
of the galaxy center with radial velocity $\geq 1000$ km/s.
Also, metal-rich gas,  recycled from circumnuclear disk, will be transported 
through the galaxy by virtue of the high-speed BAL winds.

In Figure \ref{fig:star-formation}, we show the cumulative star formation 
in the circumnuclear disk, which  is $\sim 6.5\times10^9 M_\odot$ at $t_{\rm age}=12.1$ Gyr. 
The stellar disk of the newly formed stars is of a thin geometry, 
and its size is $\lesssim150$ parsec, which coincides with the cold gaseous disk 
(cf. Figure \ref{fig:hydro-properties}). 
Recall that we have two star formation algorithms
in our model setup: one is based on the Toomre instability 
and the second on the local Jeans timescale,
with the latter allowed only in the densest gaseous zones 
with hydrogen number density $n_H>10^5~{\rm cm}^{-3}$ \citep{gan_macer_2018}.  
So, most of the star formation is confined to the innermost region of the circumnuclear disk.
In the simulation, we are able to track the age and the in-situ metallicity 
of the newly formed stars, as shown in Figure \ref{fig:star-formation}, 
the averaged age and metallicity of the new stars are 7.52 Gyr 
and $Z/Z_\odot=5.82$, respectively.

\begin{figure}[htb]
\centering
\includegraphics[width=0.475\textwidth]{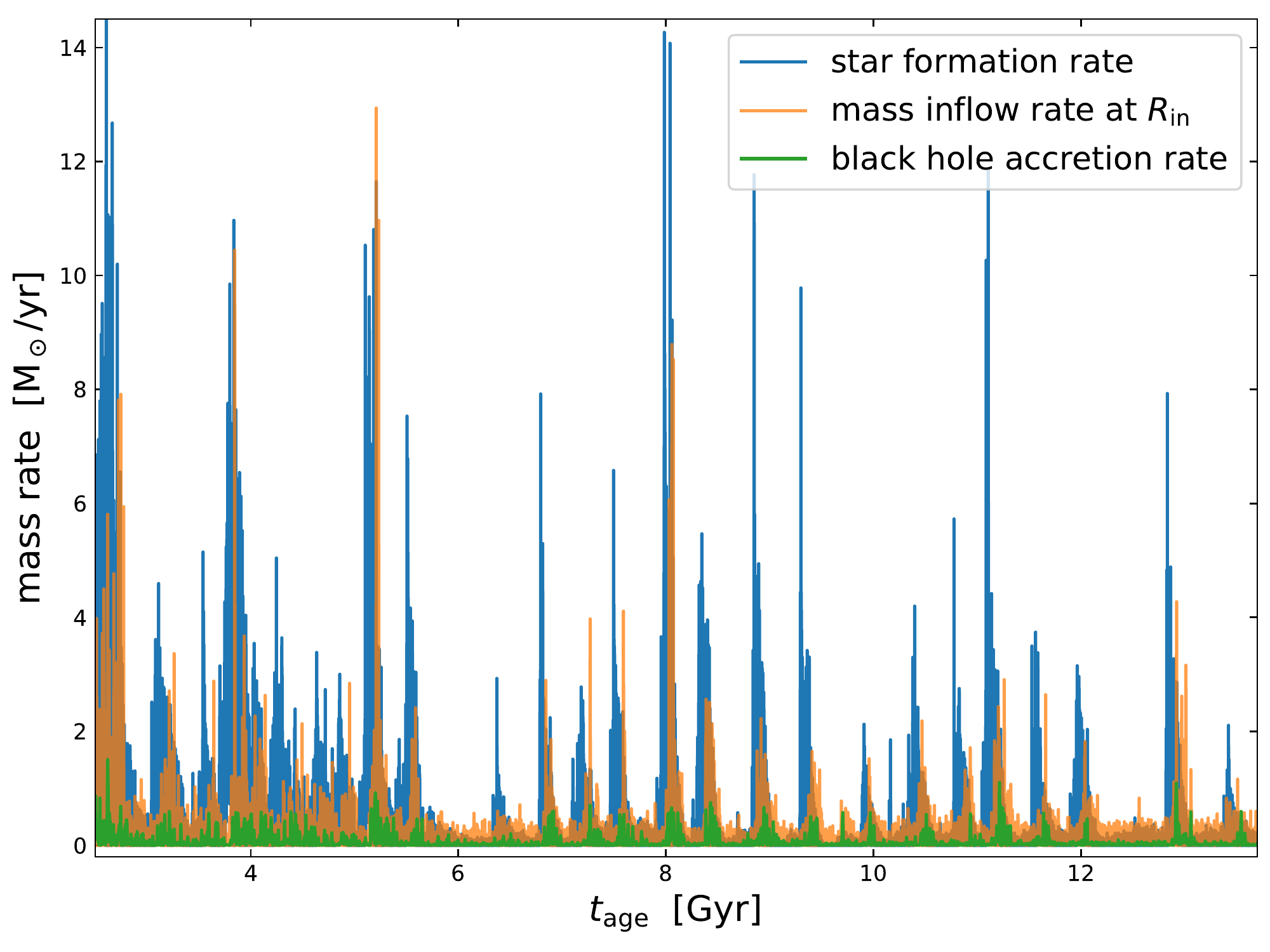}
\caption{History of the star formation and AGN activities. 
The blue,  green, and orange lines show the variations of the star formation rate, 
mass inflow rate onto the galaxy center through the inner boundary ($R_{\rm in}$), 
and the black hole accretion rate, respectively.  
As shown in the figure, black hole accretion events usually associate with star formation. \\\\
}
\label{fig:light-curves}
\centering
\includegraphics[width=0.475\textwidth]{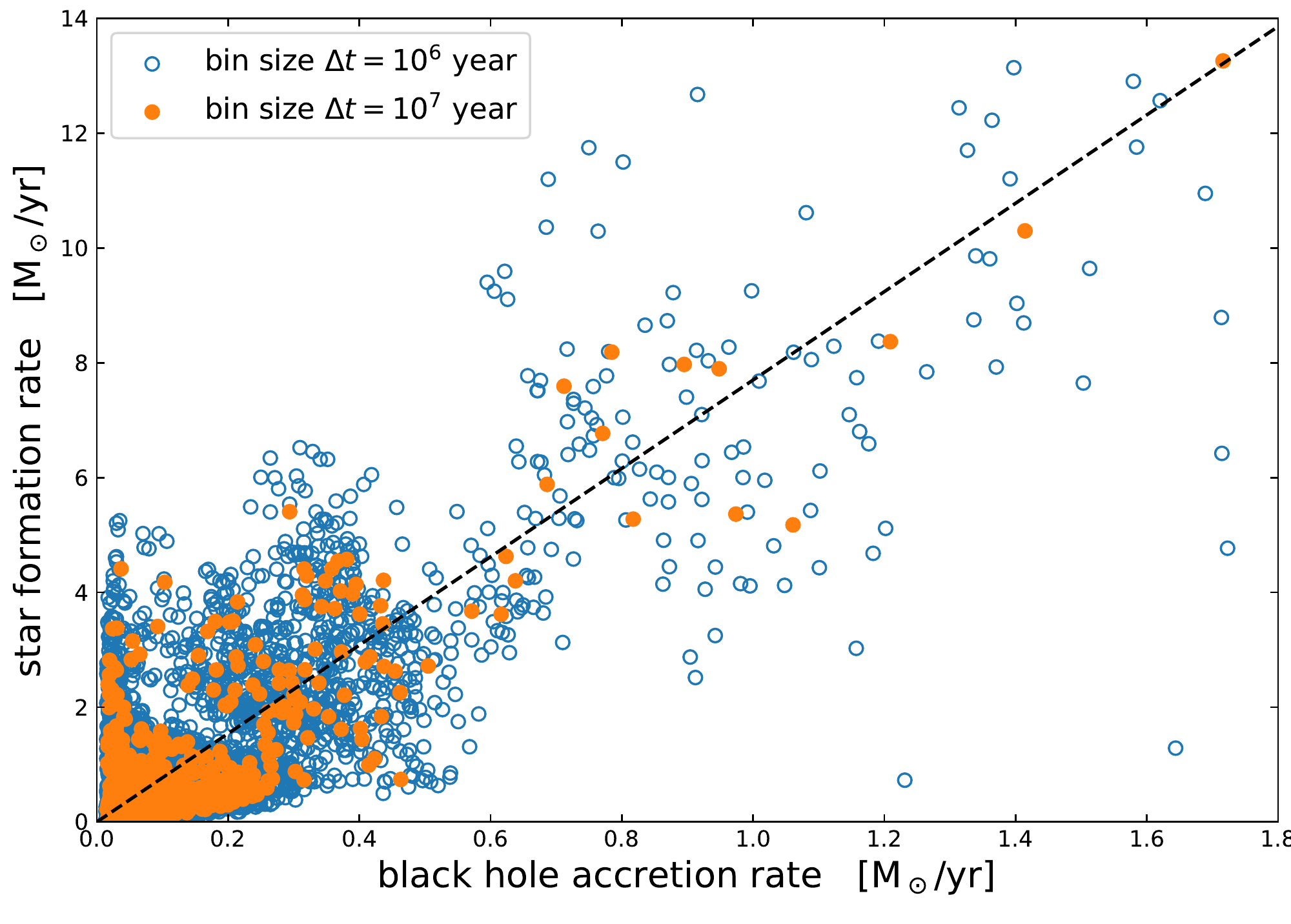}
\caption{
The scatter plot between the black hole accretion rate $\dot{M}_{\rm BH}$
and star formation rate $\dot{M}_\star^+$.  
In the figure, we bin the timing data in Figure \ref{fig:light-curves} with adjacent time intervals 
of $\Delta t = 10^6$ and $10^7$ year, respectively, 
as shown with the blue (open) and yellow (filled) cycles. 
We find a tight correlation of $\dot{M}_\star^+ \sim 7.7 \dot{M}_{\rm BH}$. 
Note that no correlation can be found before binning the data,
due to the time lags between star formation and black hole accretion.\\\\}
\label{fig:bh-sf-correlation}
\centering
\includegraphics[width=0.475\textwidth]{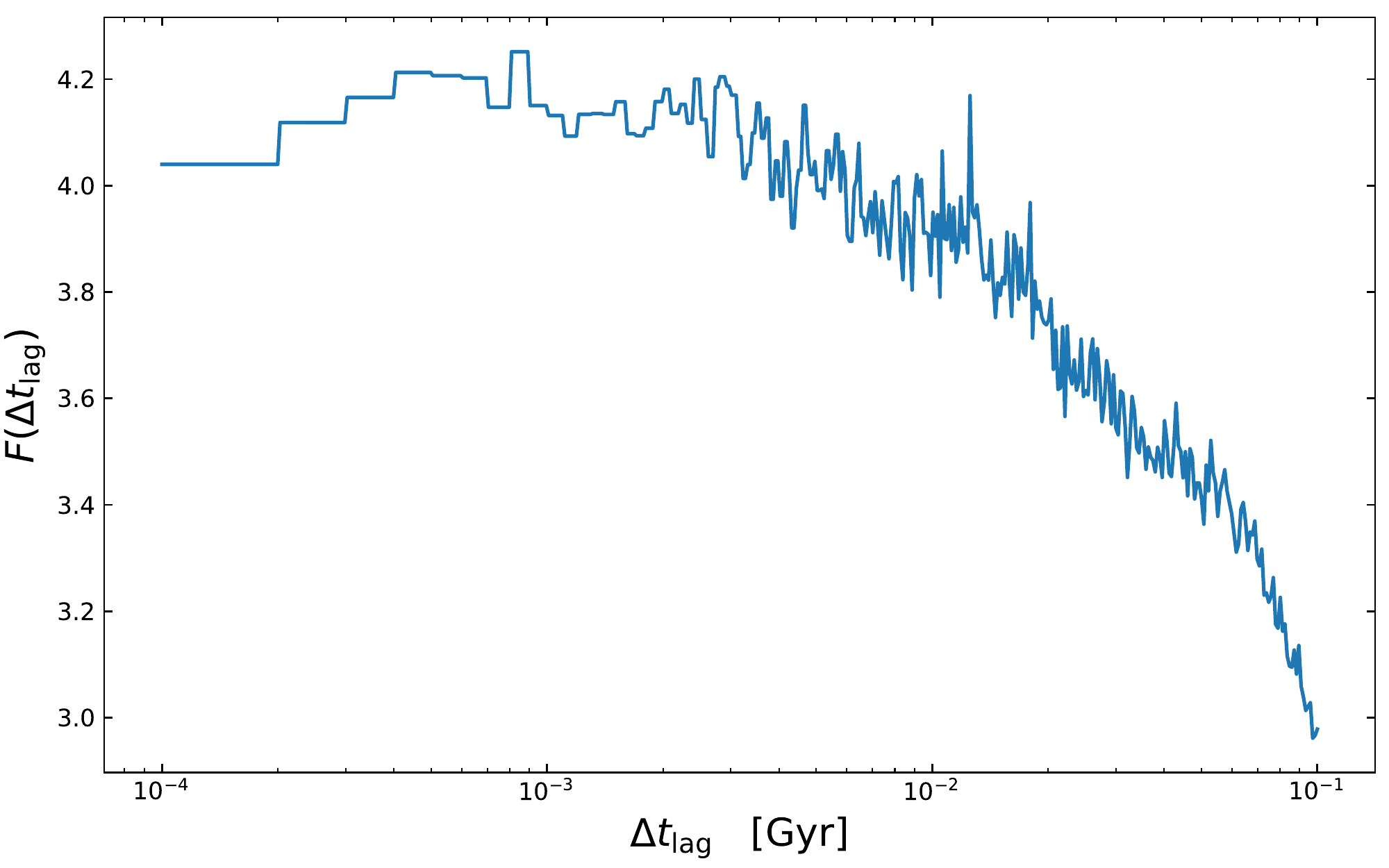}
\caption{
The time lag between star formation and black hole accretion 
(cf Equation \ref{eq:time-lag}). }
\label{fig:bh-sf-time-lag}
\end{figure}

\subsection{Star formation vs. mass accretion} \label{sec:sf-bha-correlation}
As we discussed previously, 
it is the Toomre instability that primarily triggers star formation 
and allows mass inflow simultaneously.
So it is unavoidable to expect a coincidence between AGN bursts and star formation
\citep{goodman_self-gravity_2003, Kawakatu_coevolution_2008, 
Imanishi_infrared_2011, Diamond-Stanic_relationship_2012, Esquej_nuclear_2014,
yang_black_2017, 
Izumi_circumnuclear_2016, Izumi_circumnuclear_2018}.
In Figure \ref{fig:light-curves}, we plot the history of star formation rate $\dot{M}_\star^+$ (blue line) 
and the mass inflow rates (that through the inner boundary $\dot{M}_{\rm in}$ in yellow line, 
and that onto the black hole horizon $\dot{M}_{\rm BH}$ in green line, respectively). 
As expected, the AGN activities are usually coincident roughly with star formation, 
and the instantaneous star formation rate $\dot{M}_\star^+$ is typically much larger than 
the mass inflow rate through the inner boundary $\dot{M}_{\rm in}$
(or the black hole accretion rate $\dot{M}_{\rm BH}$).
That is, star formation consumes most of the cold gas within the circumnuclear disk 
before it can fall onto the central supermassive black hole.
This also means that SNe II in the new stellar population can play a very important
role in enriching metals (see \S\ref{sec:metal-budget} for detailed analysis).

In Figure \ref{fig:bh-sf-correlation}, we present the scatter plot 
between the black hole accretion rate $\dot{M}_{\rm BH}$
and star formation rate $\dot{M}_\star^+$,  
where we bin the timing data in Figure \ref{fig:light-curves} with adjacent time intervals 
of $\Delta t = 10^6$ and $10^7$ year, respectively, 
as shown with the blue (open) and yellow (filled) cycles. 
We find a tight linear correlation of $\dot{M}_\star^+ \sim 7.7 \dot{M}_{\rm BH}$. 
Note that no correlation can be found before binning the data,
due to the time lags between star formation and black hole accretion.
To evaluate the time lags, we perform the calculation below,
\begin{equation} \label{eq:time-lag}
F(\Delta t_{\rm lag}) = \frac{\int \dot{M}_\star^+(t)\cdot \dot{M}_{\rm BH}(t+\Delta t_{\rm lag})\;dt/T}
                         {\int \dot{M}_\star^+(t)\;dt/T \cdot \int \dot{M}_{\rm BH}(t)\;dt/T}
\end{equation}
where $T=12$ Gyr is the duration of the simulation. 
The results are shown in Figure \ref{fig:bh-sf-time-lag}. 
It indicates that the time lag between star formation and black hole accretion
is about few $10^6$ year. 

\subsection{Overall chemical budget} \label{sec:metal-budget}
\begin{figure}[htb]
\centering
\includegraphics[width=0.475\textwidth]{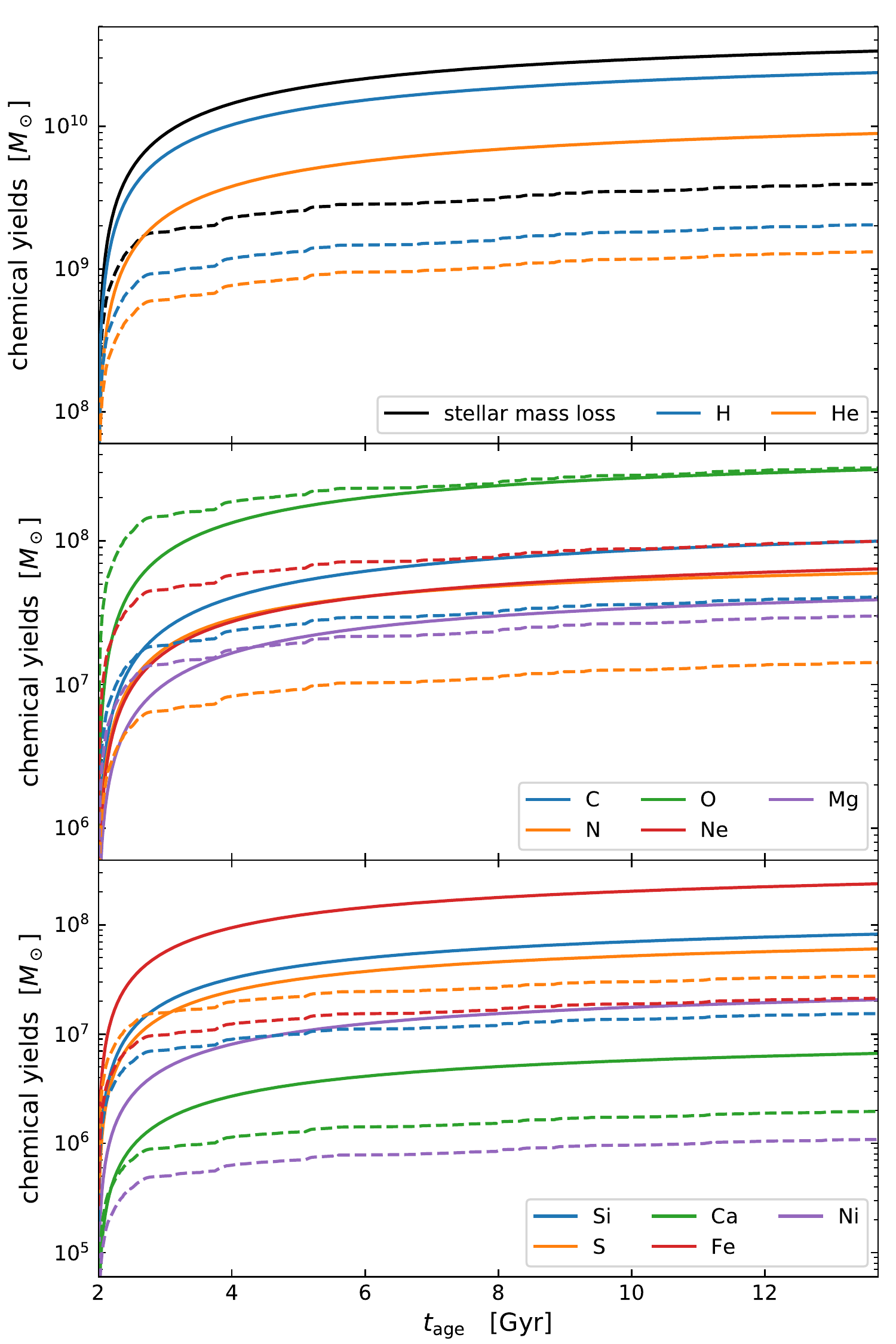}
\caption{\textcolor{black}{
Cumulative chemical yields from the initial (old) stellar population (solid lines) 
and the new stellar population (dashed lines).   
The black lines in the top panel present the cumulative stellar mass loss from the two stellar
populations, respectively, i.e., the new stellar population (black dashed line) contributes 
$\sim1/10$ of the total mass loss, while it dominates the chemical yields of Neon and
and contributes significantly to Oxygen (middle panel). 
The new stellar population also contributes a significant fraction of heavy metals 
when compared to that of the initial stellar population (bottom panel). 
}}
\label{fig:cumulative-chemical-yields}
\end{figure}

The cumulative chemical yields from the initial (old) stellar population 
and the new (young) stellar population 
are shown in Figure \ref{fig:cumulative-chemical-yields}
with the solid and dashed lines, respectively.  During the simulation, 
the old stellar population ejects in total $\sim 3.4\times10^{10} M_\odot$ 
of total mass into the ISM ($\sim 10\%M_\star$; cf Equations 
\ref{eq:stellar-mass-loss-rate} \& \ref{eq:snia-rate}). 
The cumulative star formation is $\sim 6.7\times10^9M_\odot$ in total, 
of which $58\%$ is injected into the ISM via SNe II (cf Equation \ref{eq:star-formation-imf}), 
i.e., the new stellar population contributes $\sim1/10$ of the total stellar mass loss
(see the black lines in the top panel), 
while it dominates the chemical yields of Neon and 
contributes significantly to Oxygen as shown in the middle panel. 
The new stellar population also contributes a significant fraction of heavy metals 
when compared to that of the initial stellar population as shown in the bottom panel.
Note that star formation mainly occurs in the circumnuclear disk.

With the metal sources and the flow pattern as described above, we are ready to analyze the 
metal enrichment and its transportation throughout the modeled galaxy.
As shown in Table \ref{tab:metal}, 
SN II ejecta is enriched in alpha elements (especially Ne and O);
SN Ia ejecta is of the highest mass fraction of Fe and Si; 
AGBs produce a relatively high fraction of Nitrogen.
So it is possible for us to track the stellar evolution by using the elements mentioned above.

\begin{figure*}[htb]
\centering
\includegraphics[width=0.275\textwidth]{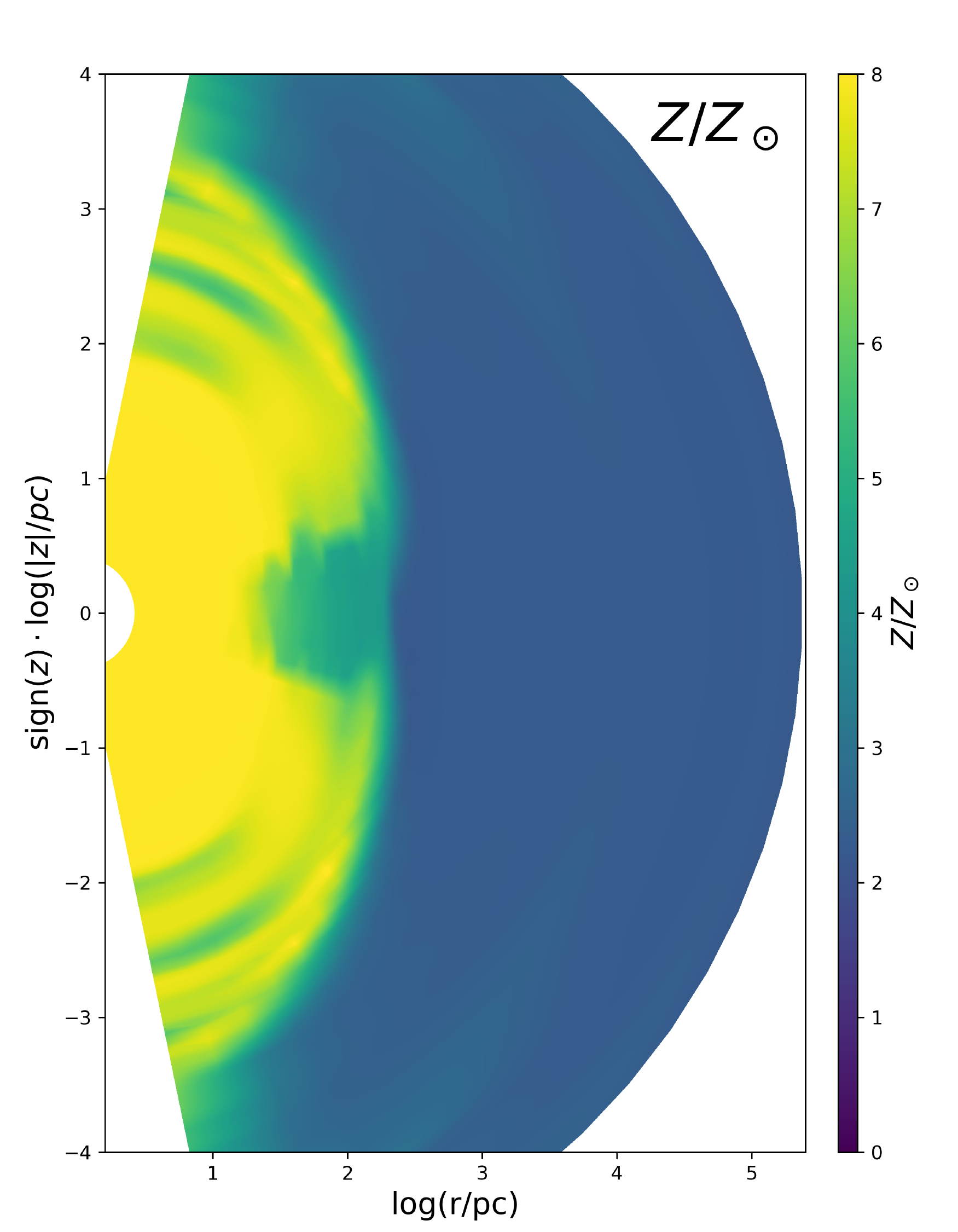}
\includegraphics[width=0.275\textwidth]{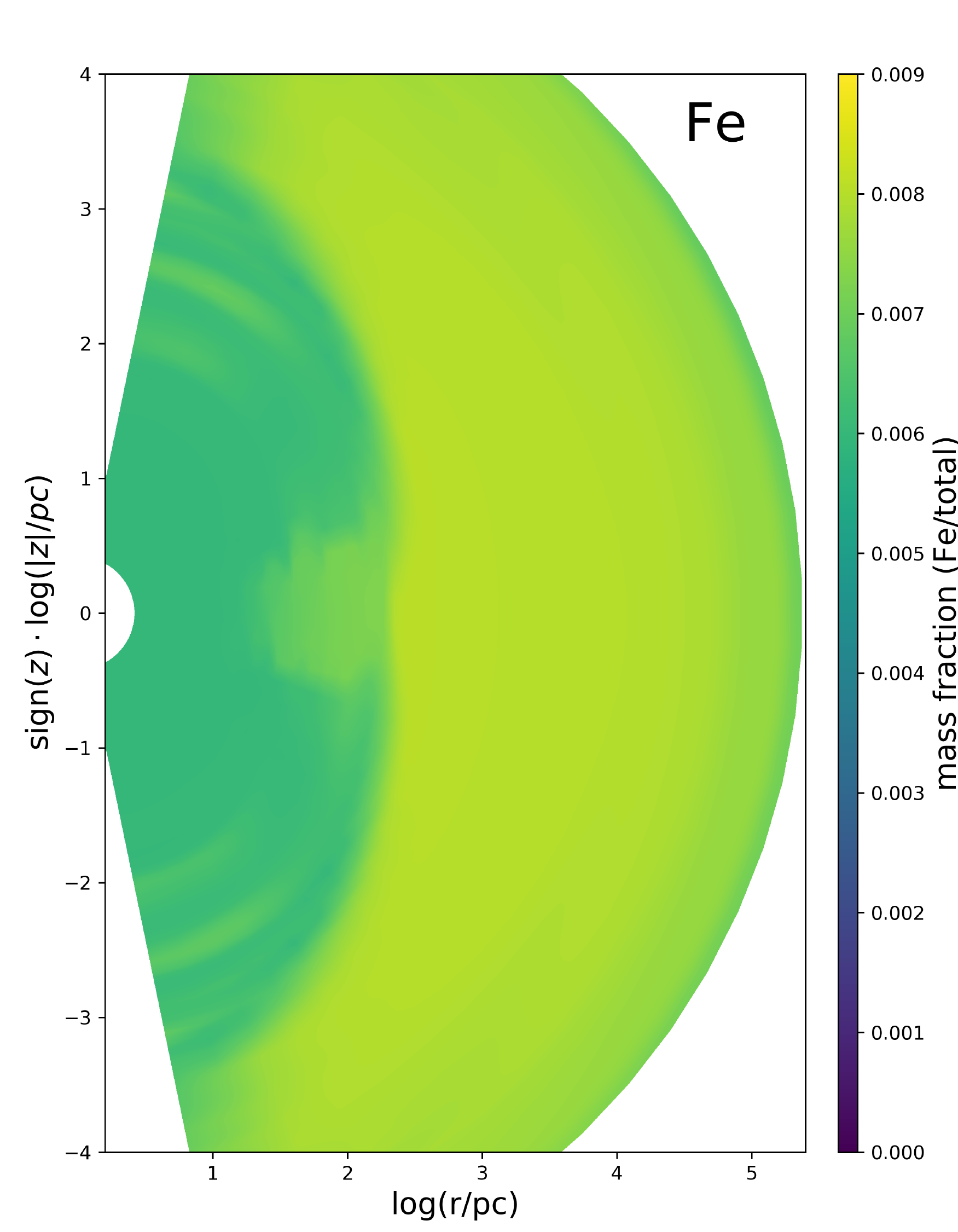}
\includegraphics[width=0.275\textwidth]{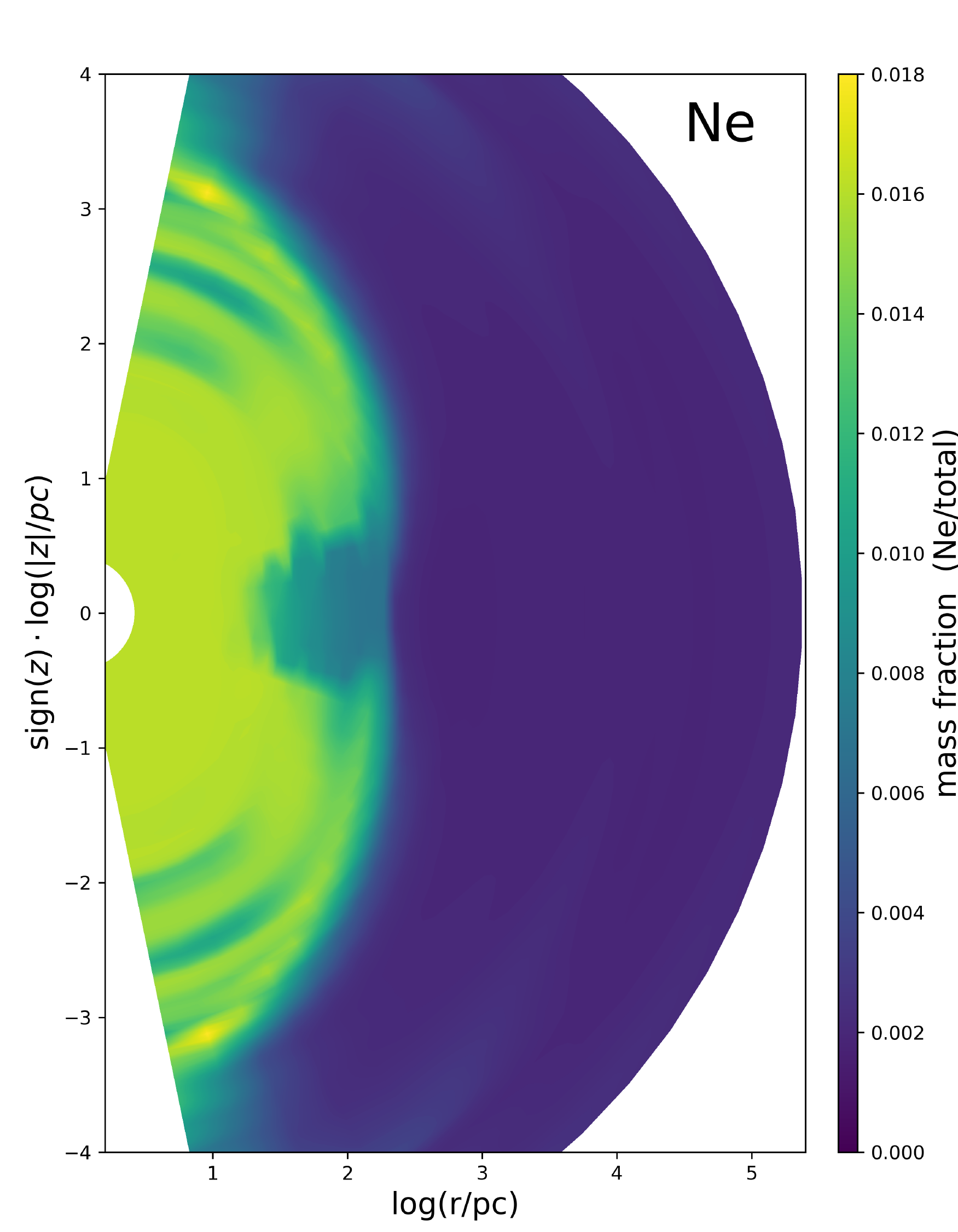}
\includegraphics[width=0.275\textwidth]{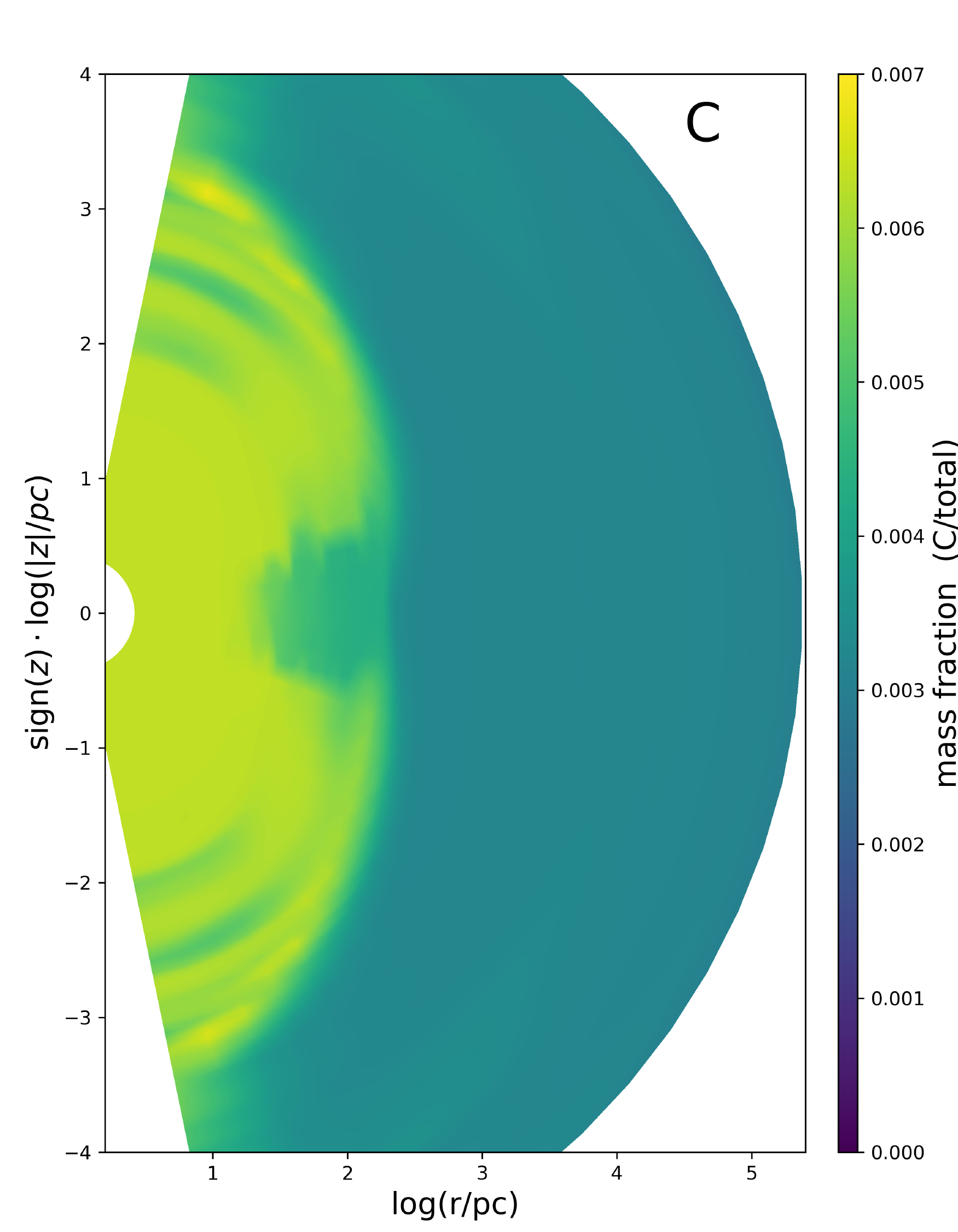}
\includegraphics[width=0.275\textwidth]{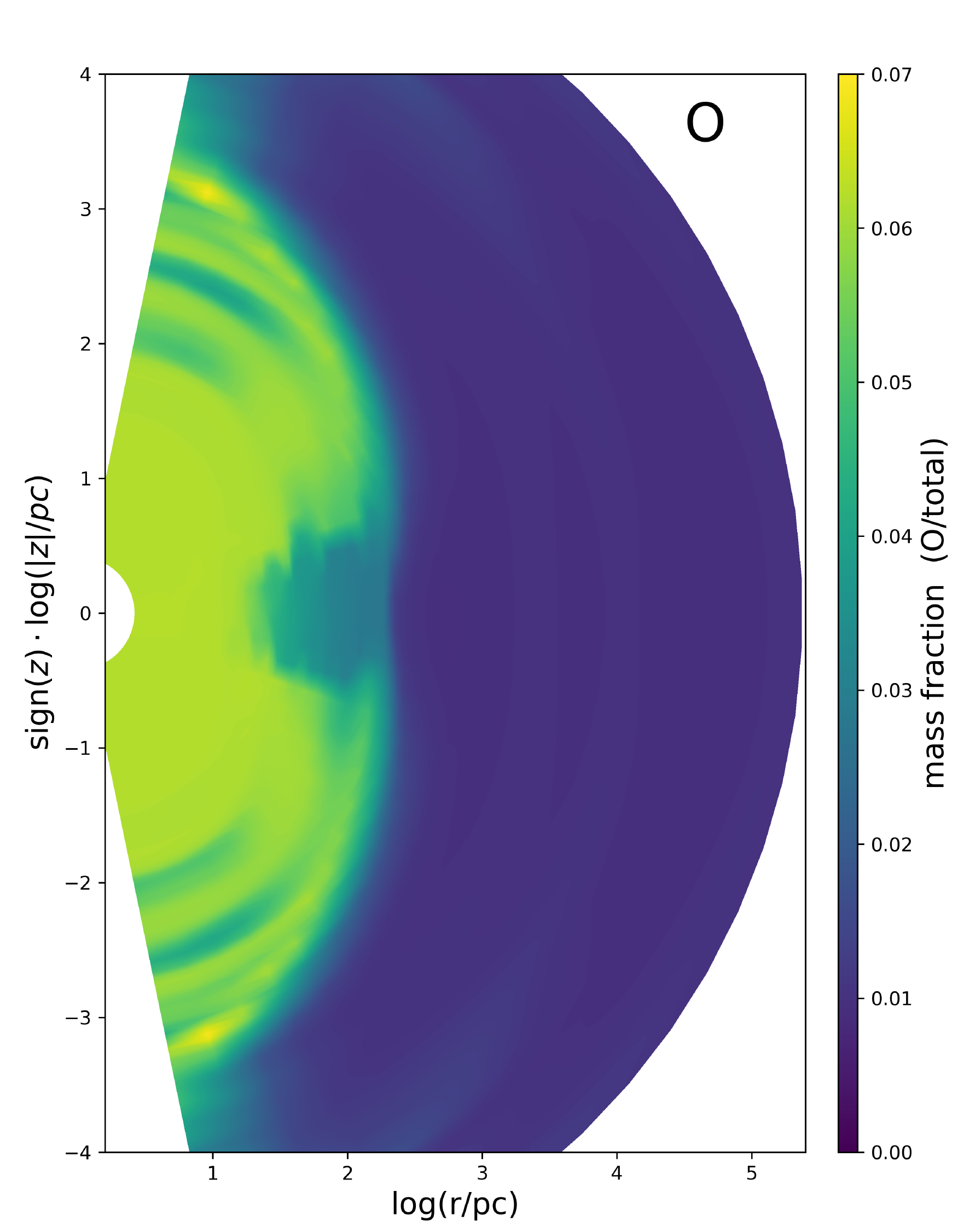}
\includegraphics[width=0.275\textwidth]{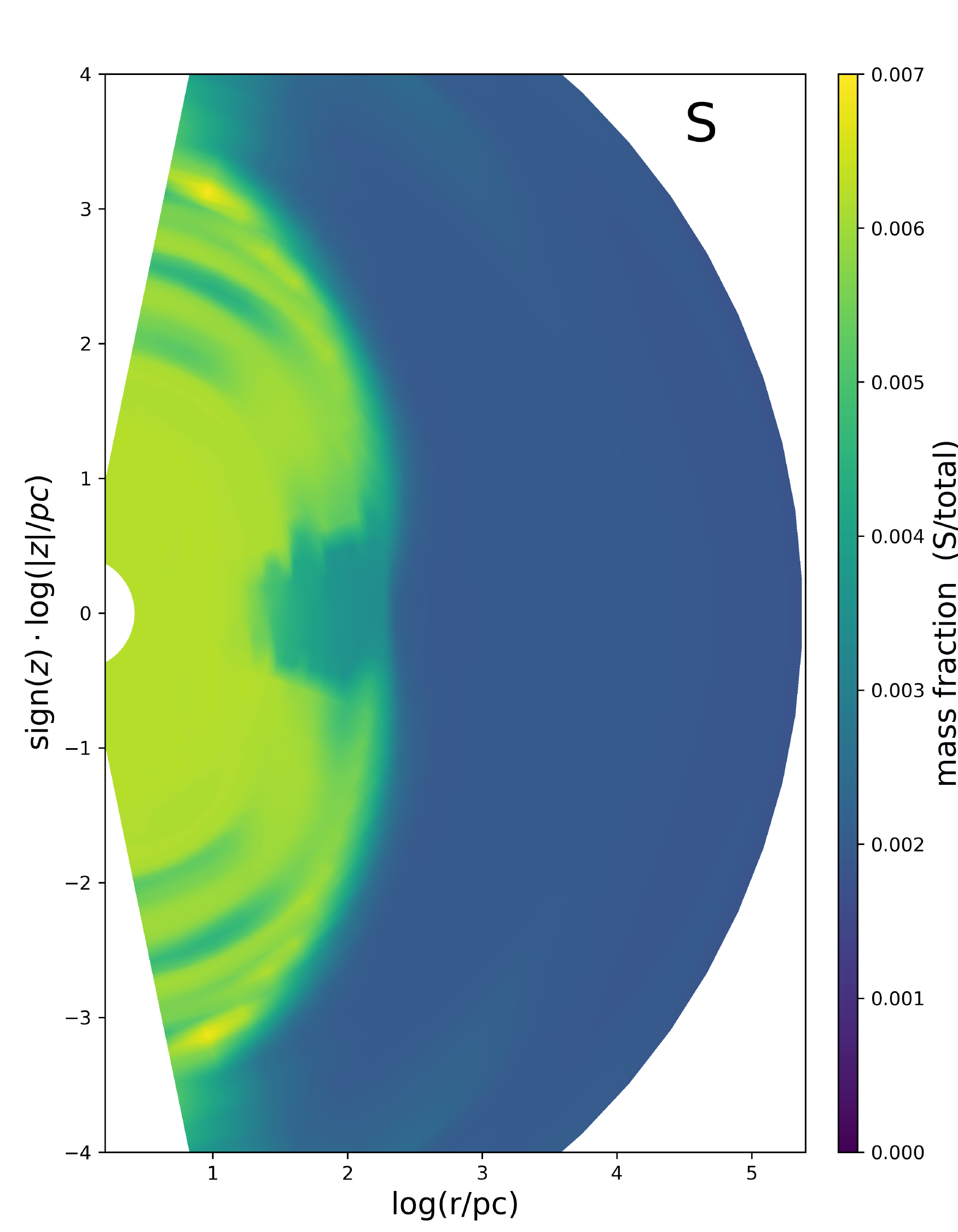}
\caption{Spatial distribution of the metallicity Z and individual chemical species (Fe, Ne, C, O, S) during the burst event at $t_{\rm age}=12.1$ Gyr. The metallicity Z is in units of the solar value ($Z_\odot=0.0134$),  and reaches a maximum value $8.04 Z_\odot$ and on average (mass-weighted) of $3.96 Z_\odot$ in the BAL winds. For the selected chemical species, the mass fractions with respect to the total local gas mass are shown in the bottom panels. Note the logarithmic radial scale.}
\label{fig:metal-distribution}
\end{figure*}

\subsection{Spatial distribution of metals} \label{sec:metal-distribution}

In Figure \ref{fig:metal-distribution}, we present the spatial distribution of the metallicity Z 
and individual chemical species (Fe, Ne, C, O, S) during the burst event at $t_{\rm age}=12.1$ Gyr.
We find that the characteristic metal abundances are different from place to place, 
accordingly, the modeled galaxy can be divided into three parts: 
(1) the circumnuclear disk, i.e., ($r=0.10$ kpc, $\theta=\pi/2$) and the surrounding region; 
(2) the BAL region, i.e., ($r=0.05$ kpc, $\theta\le\pi/6$); 
(3) the main body of the galaxy on the length scale of 10 kpc, i.e., ($r=10$ kpc, $\theta=\pi/2$).
We sample the metal abundances from the representative locations above 
and show the results in Table \ref{tab:fluid-sample}. 
The high metallicities found in the BAL region and in the cold disk 
are consistent with the SDSS observations 
(e.g., \citealt{nagao_evolution_2006,xu_evolution_2018}).

\begin{table}[ht]
\caption{Metal abundance of sampled interstellar medium$^*$}\label{tab:fluid-sample}
\begin{center}
\begin{tabular}{rccc}
\hline\hline
{  } & {hot ISM}			 	 &{cold disk} & {BAL \textcolor{black}{region}}  \\
\hline
He/H&     1.11 &     1.23 &     1.44 \\
 C/H&     1.40 &     2.04 &     3.20 \\
 N/H&     2.19 &     2.46 &     2.96 \\
 O/H&     1.76 &     5.68 &    12.8 \\
Ne/H&     1.62 &     6.49 &    15.4 \\
Mg/H&     1.79 &     4.18 &     8.55 \\
Si/H&     4.36 &     5.11 &     6.56 \\
 S/H&     6.51 &    12.7 &    23.9 \\
Ca/H&     3.48 &     4.94 &     7.64 \\
Fe/H&     6.32 &     6.03 &     5.60 \\
Ni/H&    10.3 &     9.07 &     7.00 \\
\hline
$Z/Z_\odot$&  2.34 &  4.54 &  7.96 \\
\hline \hline
\end{tabular}
\end{center}
\hangindent 0.75em	
*  The metal abundances are sampled at $t_{\rm age}=12.1$ Gyr from 
    the hot ISM ($r=10$ kpc, $\theta=\pi/2$),
    the cold circumnuclear disk ($r=0.10$ kpc, $\theta=\pi/2$), and 
    the BAL region ($r=0.05$ kpc, $\theta=\pi/6$), respectively.
    All the measurements are in the solar units as defined in Table \ref{tab:metal}.
\end{table}

As shown in top-middle panel of Figure \ref{fig:metal-distribution}, 
the main body of the galaxy shows the highest Fe abundance 
(e.g. at $r\sim10$ kpc, $\theta=\pi/2$).
As we discussed previously,  the secular stellar evolution 
of the old stellar population dominates the metal enrichment
since there is no star formation in the main body of the galaxy. 
The typical value of the metallicity Z is $\sim2.3Z_\odot$, 
and it follows the abundance pattern of AGBs + SNe Ia (cf. Table \ref{tab:fluid-sample})
as discussed at the beginning of this section.

As shown in top-left panel of Figure \ref{fig:metal-distribution}, 
the circumnuclear disk has the highest metallicity,  
with a typical value up to $\sim8 Z_\odot$ in the innermost region
and decreasing to $\sim2.3 Z_\odot$ at its outer edge. 
Obviously, SNe II play a crucial role in the metal enrichment within the circumnuclear disk.
Note that the disk here seems much more puffy than its geometry seen 
in the hydrodynamical counterpart (cf Figure \ref{fig:hydro-properties}),
which is because the massive SN II explosions contaminate the chemical composition 
above and below the disk.
We use Neon to track the contributions of SNe II to the metal enrichment 
in the upper-right panel, we can see that the spatial distribution of Neon 
follows well with the biconical structure of the BAL winds 
(cf Figure \ref{fig:hydro-properties}; 
alpha elements shown in the bottom panels of Figure \ref{fig:metal-distribution}
also share the similar patterns). 
As we mentioned, the AGN feeding process is mainly via the circumnuclear disk. 
Therefore, the BAL winds would naturally inherit the high metallicity of the circumnuclear disk, 
and spread the metal-rich gas in a fashion of biconical winds.
The metallicity of the BAL winds is up to $\sim8 Z_\odot$ and 
on average (mass-weighted) it is $\sim3.96 Z_\odot$, because it is diluted by the ISM while the wind is propagating through its host galaxy. Detailed metal composition of the BAL winds (during three representative burst events, at $t_{\rm age}=4.0, 8.1, 12.1$ Gyr, respectively) can be found in Table \ref{tab:BAL-metal},
as previously, we define the BAL winds as the components 
moving with radial velocity $\geq1000$ km/s.

We also see at the early epochs that the galaxy outskirts has extremely low metallicity
because of the dilution due to the low-metallicity CGM infall (not shown here). 
When such gas cools down and falls onto the galaxy, 
it would also dilute the metal abundance in the circumnuclear disk. 

\begin{table}[ht]
\caption{Mass-weighted metal abundance in the BAL winds$^*$}\label{tab:BAL-metal}
\begin{center}
\begin{tabular}{rccc}
\hline\hline
{  } & {Peak 1}			 	 &{Peak 2} & {Peak 3}  \\
{  } & {($t_{\rm age}=4.0$ Gyr)} &{($t_{\rm age}=8.1$ Gyr)} & {($t_{\rm age}=12.1$ Gyr)}  \\
\hline
He/H&     1.17 &     1.22 &     1.20 \\
C/H&     1.64 &     1.94 &     1.88 \\
 N/H&     2.85 &     2.57 &     2.39 \\
 O/H&     3.91 &     5.23 &     4.66 \\
Ne/H&     4.33 &     5.95 &     5.20 \\
Mg/H&     3.08 &     3.89 &     3.58 \\
Si/H&     4.16 &     4.83 &     4.95 \\
 S/H&     9.47 &    11.8 &    11.1 \\
Ca/H&     3.99 &     4.69 &     4.59 \\
Fe/H&     5.33 &     5.80 &     6.15 \\
Ni/H&     8.06 &     8.68 &     9.48 \\
\hline
$Z/Z_\odot$&     3.50 &     4.25 &     3.96 \\
\hline \hline
\end{tabular}
\end{center}
\hangindent 0.75em	
*  The measurements are made in the BAL regions
   which are selected if radial velocity $v_r\geq1000$ km/s.
   The metal fractions are weighted by the total ISM mass.
   Note that the metallicity of the BAL winds is diluted by the ISM 
   while it is propagating through its host galaxy. 
   All the measurements are in the solar units as defined in Table \ref{tab:metal}.
\end{table}

\begin{figure}[htb]
\centering
\hspace{15in}
\includegraphics[width=0.495\textwidth]{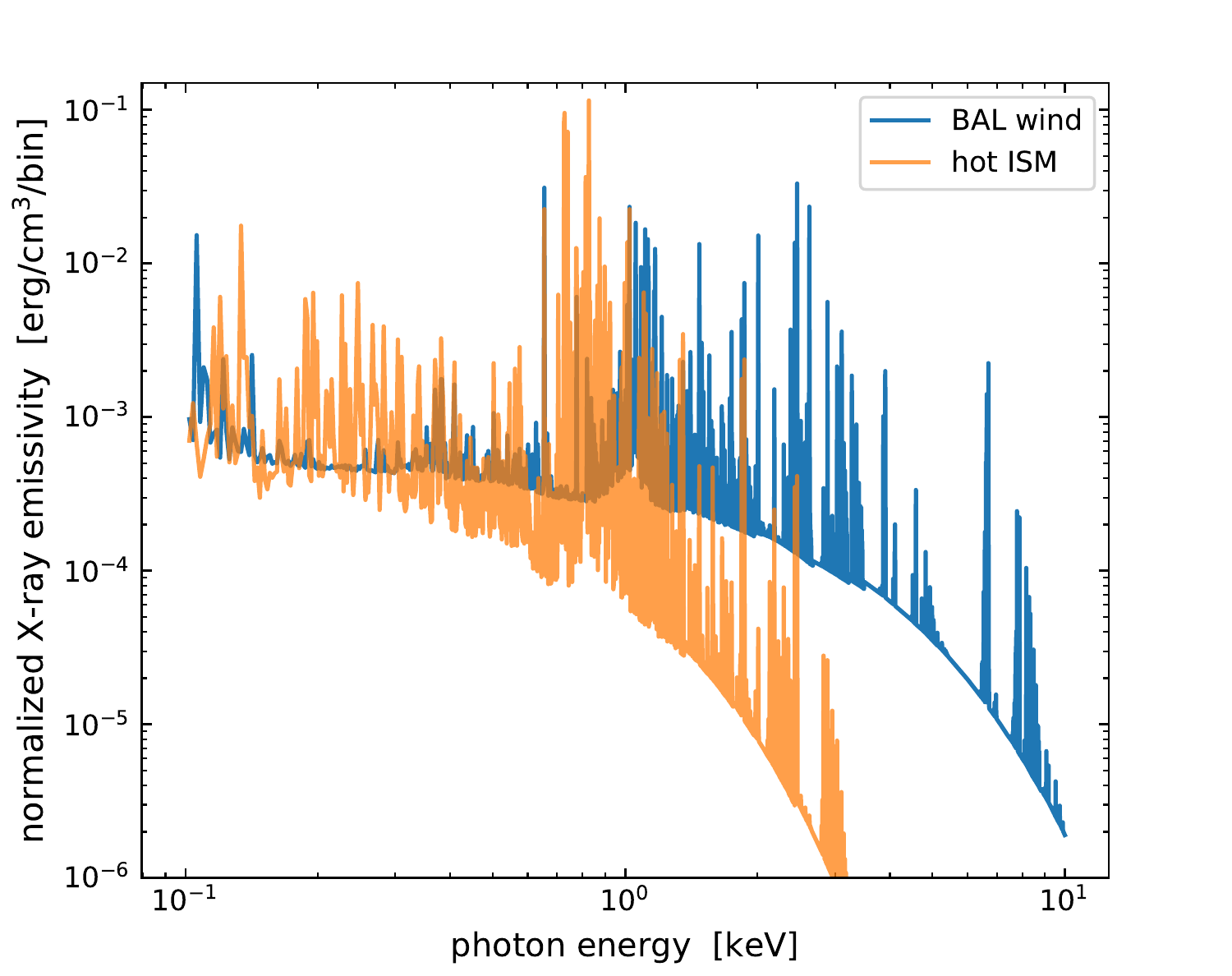}
\caption{Metallicity-dependent X-ray emissivities with the sampled metal abundances
in Table \ref{tab:fluid-sample}. The BAL sample is of a typical temperature $\sim 2\times10^7$ K,
and that of the hot ISM is $\sim 5\times10^6$ K.\\}
\label{fig:xray-spectra}
\end{figure}

\subsection{Metals seen in X-rays} \label{sec:radiative-featues}
As a result of metal enrichment, the radiative 
cooling and {\it photoionization} heating of the ISM are significantly enhanced.
We pay special attention to finding  observable features which can be used to test our model. 
In the rest of this section,
we will perform detailed analysis on the radiative features of the hot ISM, 
and illustrate the advantages of computing detailed metallicity. 
The post processing is based on the hydrodynamical data. 
The frequency-dependent radiation is calculated 
using the atomic database \texttt{ATOMDB} (version 3.0.9; 
assuming collisional equilibrium),
for example, at any given time and location, 
we retrieve the ISM properties, i.e., its density, temperature,
and the detailed abundance of the 12 chemical species.
Then, we take such output from the \texttt{MACER} simulation 
as the input of \texttt{ATOMDB},
ultimately, the latter would give us the emissivity in the energy band of interest.

\begin{figure}[htb]
\centering
\includegraphics[width=0.475\textwidth]{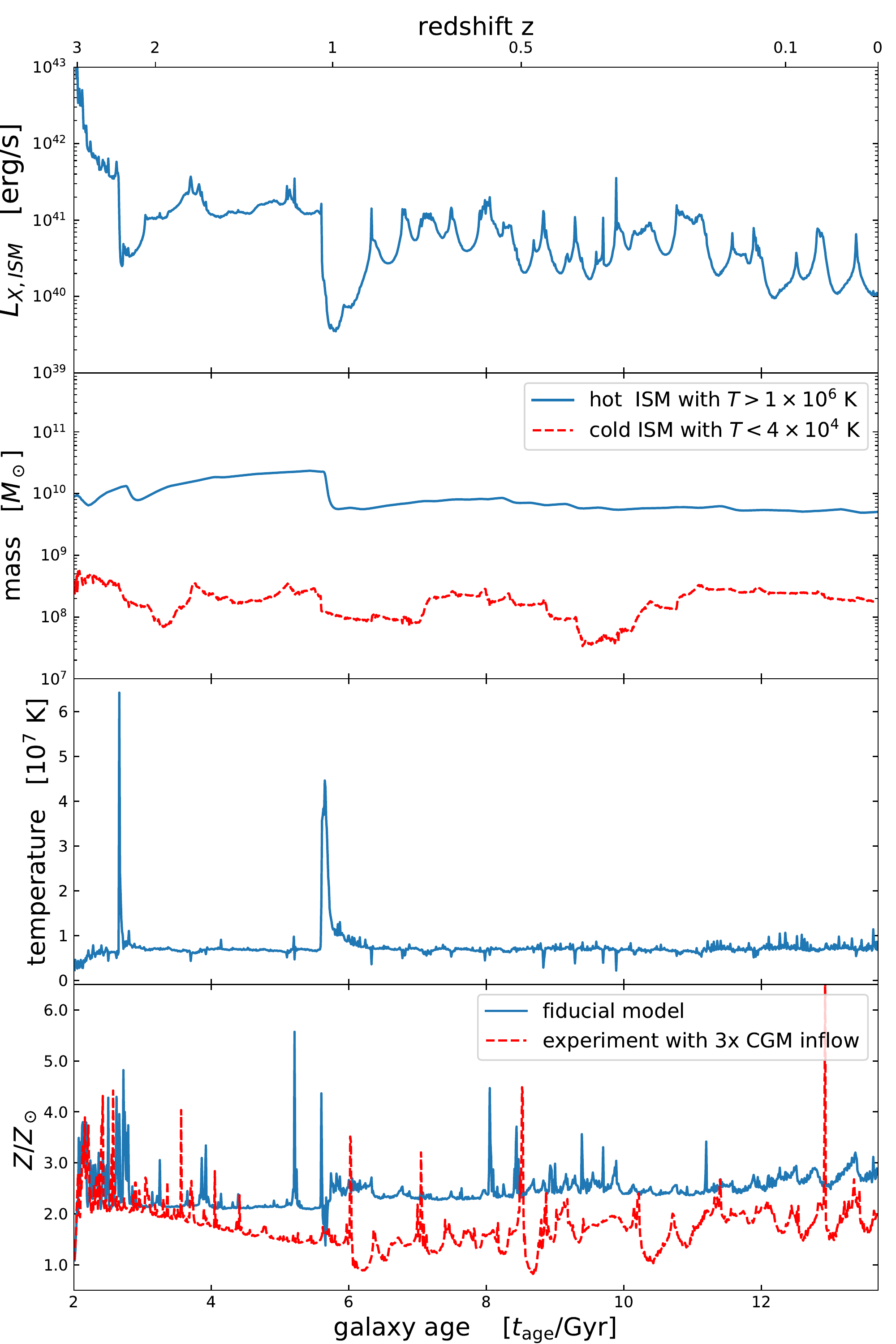}
\caption{X-ray-emission-related properties of the hot ISM. 
From top to bottom, it shows the X-ray luminosity, total ISM mass,  
X-ray-emission weighted temperature and metallicity, resepctively. 
Note that the disk region ($r<200$ parsec and $\pi/3 < \theta < 2/3\pi$) 
and the densest zones with hydrogen number density $>10^2~{\rm cm}^{-3}$ are
excluded in the calculations of the X-ray-emission weighted properties above.
The red dashed line in the bottom panel shows the X-ray-emission weighted
metallicity in an experimental run with three times higher CGM inflows 
when compared to our fiducial model in this paper.\\}
\label{fig:xray-weighted-properties}
\end{figure}

In Figure \ref{fig:xray-spectra},  we plot the representative X-ray spectra 
for the hot ISM and BAL winds as sampled in Table \ref{tab:fluid-sample}.
The BAL sample is of a typical temperature $\sim2\times10^7$ K,
and that of the hot ISM is $\sim5\times10^6$ K. 
With the metallicity-dependent emissivity, 
we are able to synthesize the radiative features of the hot gas.

In the top two panels of Figure \ref{fig:xray-weighted-properties}, 
we plot the X-ray luminosities and the mass of the hot ISM. 
We can see that  the ISM X-ray luminosity decreases
systematically from redshift z=3.2 to 0, which is mainly because of 
the decrease of gas content in the modeled galaxy.

In the lower panels of Figure \ref{fig:xray-weighted-properties}, we analyze  
the X-ray-emission weighted properties of the hot ISM. 
We average a quantity of interest $P$ in the manner below,
\begin{equation} \label{eq:xray-weighted}
     <P> \equiv \frac{\int{\; j_\nu P \; \mathrm{d} V}}{\int{\; j_\nu \; \mathrm{d} V}}. 
\end{equation}
where $j_\nu$ is the X-ray emissivity in the energy band of 0.3-8 keV. 
In the volume integrals above, we exclude the disk region 
($r<200$ parsec and $\pi/3 < \theta < 2/3\pi$) 
and the densest zones with $n_{\rm H}>10^2~{\rm cm}^{-3}$,
i.e., which corresponds approximately to a column density of $10^{23} ~{\rm atom}/{\rm cm}^2$ 
and may be optically thick to X-rays \citep{ballantyne_obscuring_2008}; 
The X-ray-emission weighted ISM temperature is presented in the second panel from bottom, 
the typical temperature is $\sim 0.7$ keV which is comparable to the virial temperature 
on the length scale of 10 kpc. 
The averaged metallicity is shown in the bottom panel, 
of which the typical value is $\sim2.5Z_\odot$,
while with Fe abundance approximately $\sim6\times$ the solar value.

\section{Conclusions and Discussions} \label{sec:conclusions}
We have developed the \texttt{MACER} code as an instrument for exploring 
the evolution of massive elliptical galaxies at high spatial resolution 
down to and within the fiducial Bondi radius, which enables us to evaluate the 
black hole feeding and feedback processes in a self-consistent way, 
and to track the coevolution between the supermassive black holes
and their host galaxies. 
In the \texttt{MACER} simulations, we paid special attention to
the internal secular stellar evolution 
which plays a crucial role in driving the galaxy evolution
as important sources of mass and energy for the ISM. 
As an increment to the \texttt{MACER} code, in this paper, 
we track the recycled mass from the stellar physics 
with its detailed chemical abundances. 
Such an increment  provides us unprecedented details of the ISM properties, 
which are testable by direct comparison with observations. 

To trace the metal enrichment, transportation and dilution processes, we solved 12 additional continuity equations dedicated to H, He, C, N, O, Ne, Mg, Si, S, Ca, Fe and Ni respectively. 
The metal yields, from AGBs and supernovae of type Ia and II,  
are calculated based on standard stellar physics. 
The chemical species are assumed to co-move once after they are injected into the ISM, 
and they are naturally mixed following the fluid motion. 

As expected, a Toomre unstable circumnuclear disk forms in the galaxy center 
with a size of $\sim150$ parsec, 
which plays a crucial role in the chemical evolution of its host galaxy
--- it is where the metals are condensed, further enriched and recycled.
The half mass radius of the new stars is roughly 20 parsec.
More specifically, the massive stars formed in the cold gas disk 
supplement a significant fraction of metals.
From Figure \ref{fig:star-formation} we see that of order $10^6$ massive stars
will be formed over the lifetime of a massive galaxy. 
In their death throes, they will produce core collapse SNe II spewing out 
alpha rich chemical products. 
As a result, the metallicity of the disk in the innermost region reaches up to $\sim8 Z_\odot$
(Figure \ref{fig:metal-distribution}). 
Such metal-rich gas will be captured by the supermassive black hole, 
and then much of it will be recycled back to its host galaxy by virtue of high-speed BAL winds. 
The latter can be readily observed. 
We found that the simulated metallicity  in the BAL winds could be up to $\sim 8 Z_\odot$, 
while that of its host galaxy is $\sim 2.3 Z_\odot$. 
The X-ray emitting hot gas is very metal enriched with a fluctuating value 
typically near $3 Z_\odot$.

Our results are well consistent with the clear correlation found observationally 
between quasar luminosities and their nuclear gas metallicity 
\citep{warner_relation_2003, nagao_evolution_2006, 
matsuoka_mass-metallicity_2011, xu_evolution_2018}.
\citet{xu_evolution_2018} analyzed the metallicity of quasar broad line region (BLR)
with a large SDSS sample. They found that the metallicity in the most massive galaxies
is $\sim 2 Z_\odot$, while the BLR metallicity is $0.3 \sim 1.0$ dex larger 
than their host galaxies and it does not evolve with cosmic time.
It indicates that the metal enrichment is due to recent star formation
rather than secular stellar evolution, as we found in this paper.
In the \texttt{MACER} simulation, the spatial resolution is close to the BLR length scale,
we also found clear correlation between star formation and AGN bursts
(but lagged in time by roughly $10^6$ years),
and thus significant metal enrichment in the galaxy center.
The ratio of star formation in the circumnuclear disk to 
accretion towards to the central supermassive black hole is 2.3, 
and of that amount the bulk (62\%) is blown out in the BAL winds
with a fraction of 38\% finally accreted onto the black hole.

Our simulated metallicity, both for the innermost central region 
and for the main body of the modeled galaxy, is larger than 
derived from X-ray observations 
(e.g., see \citealt{humphrey_chandra_2006}, \citealt{mernier_radial_2017} for the abundance of the hot ISM of early-type galaxies). However, there are a number of uncertainties when deriving abundances from the X-ray spectra (see \citealt{kim_metal_2012} for a review). Also, 
dust depletion of metals could be a significant factor 
(e.g., see \citealt{lakhchaura_possible_2018} for a recent analysis). Additionally we assumed
a metallicity of $1.5Z_\odot$ which seemed reasonable given that
the velocity dispersion of our test galaxy is roughly double that in the MW spheroid.
However, as noted earlier, \cite{conroy_early-type_2014} would give our test galaxy 
a Fe abundance (in the old stars) slightly less than the solar value.

One important caveat should be noted in our discussion of the expected
chemical abundances in the X-ray emitting gas. We allowed for CGM infall
of metal poor gas, but the amount of infall in our fiducial simulation was
only $1/12$ of the stellar mass of the initial galaxy, and even somewhat less than
that ($\sim1/20$) when we corrected for starting at $2$ Gyr. This could be too low
an estimate and, in any case, will vary from case to case depending on the
detailed cosmic environment. To check on the sensitivity to this component
we have run a case with three times higher CGM inflow ($1/4$ of the initial 
stellar mass corrected to $\sim1/7$ of the stellar mass at $t = 2$ Gyr). In this test run
the CGM inflow dilutes the metallicity enrichment to such an extent that the
final X-ray-emission weighted metallicity 
drops to $\sim1.5$ times solar by roughly a factor
of two from our fiducial run. Thus the expected metallicity of the hot Xray
emitting gas has an almost factor of two uncertainty depending on the 
cosmic environment --- an uncertainty that can only be addressed by simulations
that are both cosmologically correct and have very high internal spatial
resolution.
\textcolor{black}{We leave it to our future work to explore the consequences of higher 
CGM inflow (at 1/7 of the initial stellar mass) as indicated by cosmological simulations.}

The massive stars formed in the Toomre unstable central cold gas disk 
should have dramatic observable consequences, besides the metal enrichment.
For example, the SN II going off in the central disk will also produce
a copious X-ray output. 
However, those SN remnants occur in such dense regions 
that even the X-ray may not be observable
except from the ``runaway'' star explosions and these are not 
allowed for in the current simulations.

In addition, the (million) neutron star and black hole condensed remnants of the exploded
early type stars must be considered. If they survive in the discs, then 
accretion onto them should have major consequences and they would
be dramatically visible in the ``E + A'' phases when the gas discs have
largely dissipated. But there is another phenomenon which may intervene.
The embedded condensed remnants will interact with the dense gas
disk  and be dragged into the central supermassive black hole 
in a manor labeled Type I migration first discussed by 
\citeauthor{goldreich_disk-satellite_1980} 
(\citeyear{goldreich_disk-satellite_1980}; which is important in planet formation). 
Using the Equation 70 of \citet{tanaka_three-dimensional_2002}, 
we estimate that the migration time is comparable to the Hubble time especially for
$10M_\odot$ black holes. Using Type I migration underestimates the migration rate
since it does not allow for the fact that the discs are so dense that in places
they are marginally Toomre unstable. This predictable set of processes 
would produce high mass ratio black hole captures 
(a.k.a. ``Extreme Mass Ratio Inspiral", or in short EMRI) which, 
while undetectable by LIGO, 
would be candidates for LISA detection (e.g., \citealt{babak_science_2017}). 
A zeroth order estimate of the rate gives $10^{2.5}$  events/year within 1000 mpc.

In this paper, we did not consider the effect of runaway stars 
in the star-forming circumnuclear disk, which could  be 
important in spreading metals in the galaxy center. 
We also did not  yet consider the dust effects in depleting metals and obscuring radiation, 
which could alter the radiative features of the circumnuclear disk significantly.
In the near future, we will introduce the above effects into 
the \texttt{MACER} simulations, aiming to interpret/predict 
the observed/observable radiative features of the ``E+A'' phenomena. 


\section*{Acknowledgement}
We thank Kengo Tomida for helping us to add tracers into the \texttt{Athena++} code.
\textcolor{black}{We thank Gregory S. Novak for sharing the first 2D version of 
the \texttt{MACER} code in 2011, which was using \texttt{ZEUSMP/1.5}.}
We thank Jeremy Goodman, James Stone, Pieter van Dokkum,
Nadia Zakamska, Takayuki Saitoh,Tuguldur Sukhboldfor and Charlie Conroy 
for useful discussions. 
ZG is supported in part by the Natural Science Foundation of Shanghai 
(grant 18ZR1447200) and by the Chinese Academy of Sciences 
via the visiting scholar program. 
This work was done during ZG's visit to the department of astronomy 
in Columbia University. 
We acknowledge computing resources from Columbia University's 
Shared Research Computing Facility project,
  which is supported by NIH Research Facility Improvement 
  Grant 1G20RR030893-01, and associated funds from 
  the New York State Empire State Development, Division of Science Technology 
  and Innovation (NYSTAR) Contract C090171, both awarded April 15, 2010. 
Some of the simulations presented
were performed with the computing resources made available via the 
Princeton Institute for Computational Science and Engineering.


\bibliography{2018Paper2}  
\bibliographystyle{apj}
\end{document}